\documentclass[superscriptaddress,prl,english,floatfix,twocolumn,10pt]{revtex4-2}
\usepackage{graphicx}
\usepackage{float}
\usepackage{natbib}
\usepackage{amssymb}
\usepackage{amsmath}
\usepackage{mathtools}
\usepackage[utf8]{inputenc}
\usepackage{braket}
\usepackage{subfigure}
\usepackage{pgfplots}
\usepackage{csquotes}
\usepackage{hhline}
\usepackage{xr}
\usepackage{dcolumn}
\usepackage{tabularx}
\usepackage[colorlinks=true,linkcolor=blue,citecolor=blue,urlcolor=blue]{hyperref}
\usepackage{longtable}
\usepackage{listings}
\usepackage{color}
\usepackage[export]{adjustbox}
\newcolumntype{C}{>{\centering\arraybackslash}X}

\begin{document}
	
	\title{\textbf{Extended unitarity and absence of skin effect in periodically driven systems}}
	\author{Aditi Chakrabarty}
	\email{aditichakrabarty030@gmail.com}
	\affiliation{Department of Physics and Astronomy, National Institute of Technology, Rourkela, Odisha-769008, India}
	\author{Sanjoy Datta}
	\email{dattas@nitrkl.ac.in}
	\affiliation{Department of Physics and Astronomy, National Institute of Technology, Rourkela, Odisha-769008, India}
	\date{\today}
	
	\begin{abstract}		
		One of the most striking features of non-Hermitian quasiperiodic systems with arbitrarily small asymmetry in the hopping amplitudes and open boundaries is the accumulation of all the bulk eigenstates at one of the edges of the system, termed in literature as the \textit{skin effect}, below a critical strength of the potential.
		In this Letter, we uncover that a time-periodic drive in such systems can eliminate the SE up to a finite strength of this asymmetry. Remarkably, the critical value for the onset of SE is independent of the driving frequency and approaches to the static behavior in the thermodynamic limit. We find that the absence of SE is intricately linked to the emergence of \textit{extended unitarity} in the delocalized phase, providing dynamical stability to the system. Interestingly, under periodic boundary condition, our non-Hermitian system can be mapped to a Hermitian analogue in the large driving frequency limit that leads to the extended unitarity irrespective of the hopping asymmetry and the strength of the quasiperiodic potential, in stark contrast to the static limit. Additionally, we numerically verify that this behavior persists
		Based on our findings, we propose a possible experimental realization of our driven system, which could be  used as a switch to control the light funneling mechanism.
	\end{abstract}
	
	\maketitle
	
	\textit{Introduction-} 
	In recent years, non-Hermitian Hamiltonians have garnered widespread attention due to their ability to accurately mimic experimental procedures involving interactions with the environment. They also offer possibilities for exotic phases of quantum matter that are typically absent in their Hermitian counterparts. Unlike Hermitian systems, the reality of eigenenergies and stable unitary dynamics is guaranteed only in specific classes of non-Hermitian Hamiltonians that possess either $\mathcal{PT}$-symmetry \cite{Bender_1998,Bender_2002} or pseudo-Hermiticity \cite{Mostafazadeh,Kawabata}.\\
	\indent
	Among the non-Hermitian systems, the paradigmatic Hatano-Nelson (HN) Hamiltonian with asymmetric hopping amplitudes \cite{HatanoNelson1996,HatanoNelson1998} stands out as another class of non-Hermitian systems of particular interest. This interest stems from a unique phenomenon known as the \textit{skin effect} (SE), which refers to the exponential localization of all bulk eigenstates at the edges of a lattice with an open boundary \cite{Lee,Lee_2019,Sato}. Interestingly, the spectral behavior in these systems is drastically sensitive to the choice of boundary conditions \cite{Xiong_2018}. Moreover, recent studies have illustrated the existence of SE accompanying a delocalization-localization (DL) phase transition in such non-Hermitian quasicrystals \cite{Longhi}.\\
	\indent
	In parallel, the exploration of a diversified range of periodically driven systems has been triggered due to the presence of rich and intriguing features typically absent in their temporally static counterparts \cite{Lignier,Ivanov,Poli,Chen,Bitter}. The underlying quantum mechanics and the general understanding of such systems influenced by some external time-periodic drive (generally known as Floquet systems) have gained interest in recent years due to their applications in ultrafast spintronics \cite{Oka,Losada}, quantum optics \cite{Eckhardt}, ultra-cold atomic systems \cite{Meinert,Sandholzer}, and trapped ions \cite{Ran}.
	On the other hand, the interplay between periodic driving and non-Hermiticity has led to several novel findings in recent years \cite{Valle,Liu_2022,Zhou_2021,Huang,Koutserimpas,Hockendorf,Banerjee,Sengupta}. It has been demonstrated that a time-periodic drive can induce SE in systems with an on-site loss \cite{Ke}. Additionally, Floquet engineering has been utilized to control the direction of edge modes \cite{Li}. Furthermore, it has been pointed out that a time-periodic drive can stabilize the dynamics of a two-level non-Hermitian Rabi model \cite{Gong}, where the Floquet quasienergies turn out to be real, leading to \textit{extended unitarity} (EU) at the end of each complete driving period.\\
	\indent
	In this Letter, we unveil for the first time that a time-periodic drive can lead to the EU condition in HN quasicrystals, with a concurrent disappearance of the SE up to a finite strength of the non-reciprocity in the hopping amplitudes. In the limit of high frequency in the drive, we illustrate analytically that such a time-periodic modulation reduces the non-Hermitian system to a Hermitian equivalent counterpart under the periodic boundary condition (PBC), giving rise to completely real Floquet quasienergies irrespective of the strength of the potential and the asymmetry in the hopping amplitude, at each stroboscopic time period. This is in stark contrast to static HN quasicrystals, where the energy spectrum undergoes a complex to real transition at a critical value of the potential, although both these systems manifest a DL transition. Surprisingly, however, under an open boundary condition (OBC), the EU condition persists only up to a critical strength of the non-reciprocity in the delocalized regime. Interestingly, we demonstrate that the emergence of EU destroys the SE. Moreover, contrary to the static limit, where the energies are real in the SE phase, we find that in time-periodic systems, it can exist when the Floquet quasienergy spectrum is complex. These counter-intuitive results on the relationship between the SE and the eigenenergies completely alter our understanding of conventional HN systems.\\
	\indent
	In practice, from the perspective of experimentalists, it is noteworthy that the dynamics in such systems has found wide realizations in photonics and in quantum systems by enhancing the extent of optical sensing \cite{McDonald,Carlo}. Additionally, the dynamics in a Hamiltonian with asymmetric hopping has recently been studied experimentally in photonic lattices to focus an incident light at a desired location, irrespective of the point of excitation, a phenomenon termed as light funneling \cite{Weidemann}. In this Letter, we propose an experimental setup that can exploit our findings to tune the light funneling effect.\\
	\indent	\textit{Model and methods-} We consider a time-dependent version of the single-particle quasiperiodic non-Hermitian Hamiltonian \cite{Jiang,Longhi} defined as,
	\begin{eqnarray}
		\mathcal{H}(t)= \displaystyle\sum_{n=1}^{N-1} (Je^{h\text{cos}(\omega t)} c^\dag_{n+1} c_{n}
		+Je^{-h\text{cos}(\omega t)} c^\dag_{n} c_{n+1})\nonumber \\
		~~~~~~~~+\sum_{n=1}^N V \text{cos} (2\pi\alpha n) c^\dag_{n} c_{n},~~~~~~~~~~~~~~~~
		\label{Eq:Hamiltonian}
	\end{eqnarray}
	 where $c^\dag_{n}$ and $c_{n}$ denote the fermionic creation and annihilation operators, $N$ represents the number of sites in the lattice, where $n$ is the site index. The lattice size $L=Na$, where $a$ is the lattice period of translation (considered to be 1 in arb. units). In the prototypical static HN Hamiltonian, the asymmetric hopping of the fermions towards the left and the right is incorporated in the tight-binding notation using an imaginary magnetic vector potential by the terms $e^{-h}$ and $e^{h}$ respectively. In this Letter, we consider continuous temporally cosine modulated non-reciprocal hopping amplitudes as indicated in the first term of the above Hamiltonian. The second term characterizes the on-site quasiperiodic potential of the Aubry-Andr{\'e}-Harper (AAH) type, where $\alpha$ defines the incommensurability, set as $(\sqrt{5}-1)/2$ throughout this work. It is evident that after a stroboscopic period $T$, $\mathcal{H}(t)=\mathcal{H}(t+T)$.\\
	 \indent For the time-periodic Hamiltonian as given in Eq.~\ref{Eq:Hamiltonian}, the Floquet theory has been instrumental in determining the states after a time $T$. According to the Floquet theory, the Floquet propagator for one complete period and an initial time $t_0=0$ is defined as,
	 \begin{eqnarray}
	 	U(T,0)=\mathcal{T}e^{-(i/\hbar)\int_{0}^{T}\mathcal{H}(t)dt}=e^{-i\mathcal{H}_FT/\hbar},
	 	\label{Eq:Floquet_operator}
	 \end{eqnarray}
	 where $\mathcal{T}$ takes care of the time-ordering of the Hamiltonians at different instants of time. In the above equation, $\mathcal{H}_F$ is the Floquet Hamiltonian, whose eigenstates can be obtained by the exact diagonalization of $U(T,0)$.
	 The reduced Planck's constant ($\hbar$) is considered to be of unit magnitude throughout this work. In general,
	 $U(T,0)$ is non-unitary when the elemental Hamiltonian is non-Hermitian, and can be constructed using the biorthogonal formalism \cite{Brody}.\\
	 \begin{figure}[]
	 	\begin{tabular}{p{\linewidth}c}
	 		\centering
	 		\includegraphics[width=0.244\textwidth,height=0.225\textwidth]{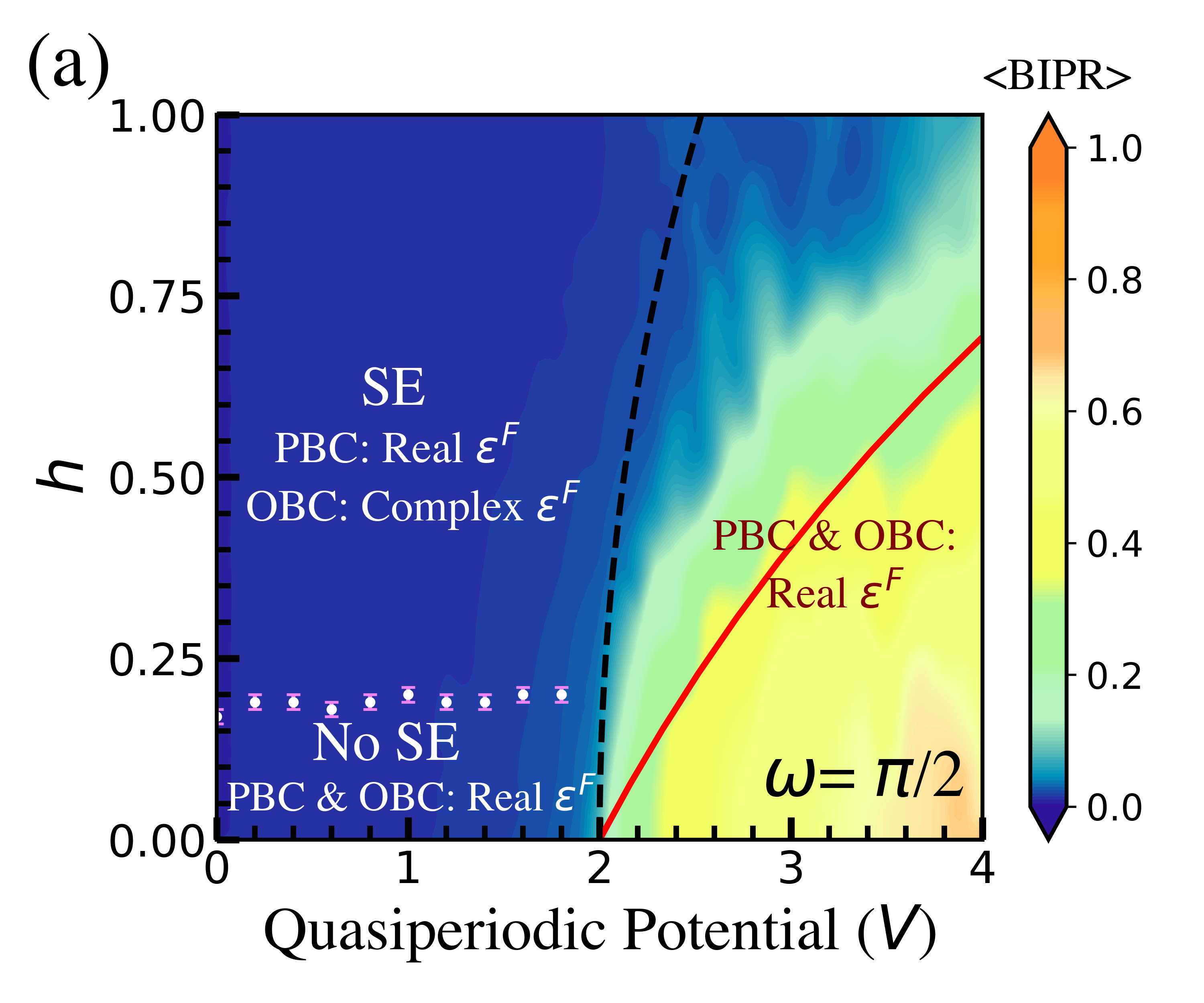}\hspace{-0.2cm}
	 		\includegraphics[width=0.244\textwidth,height=0.225\textwidth]{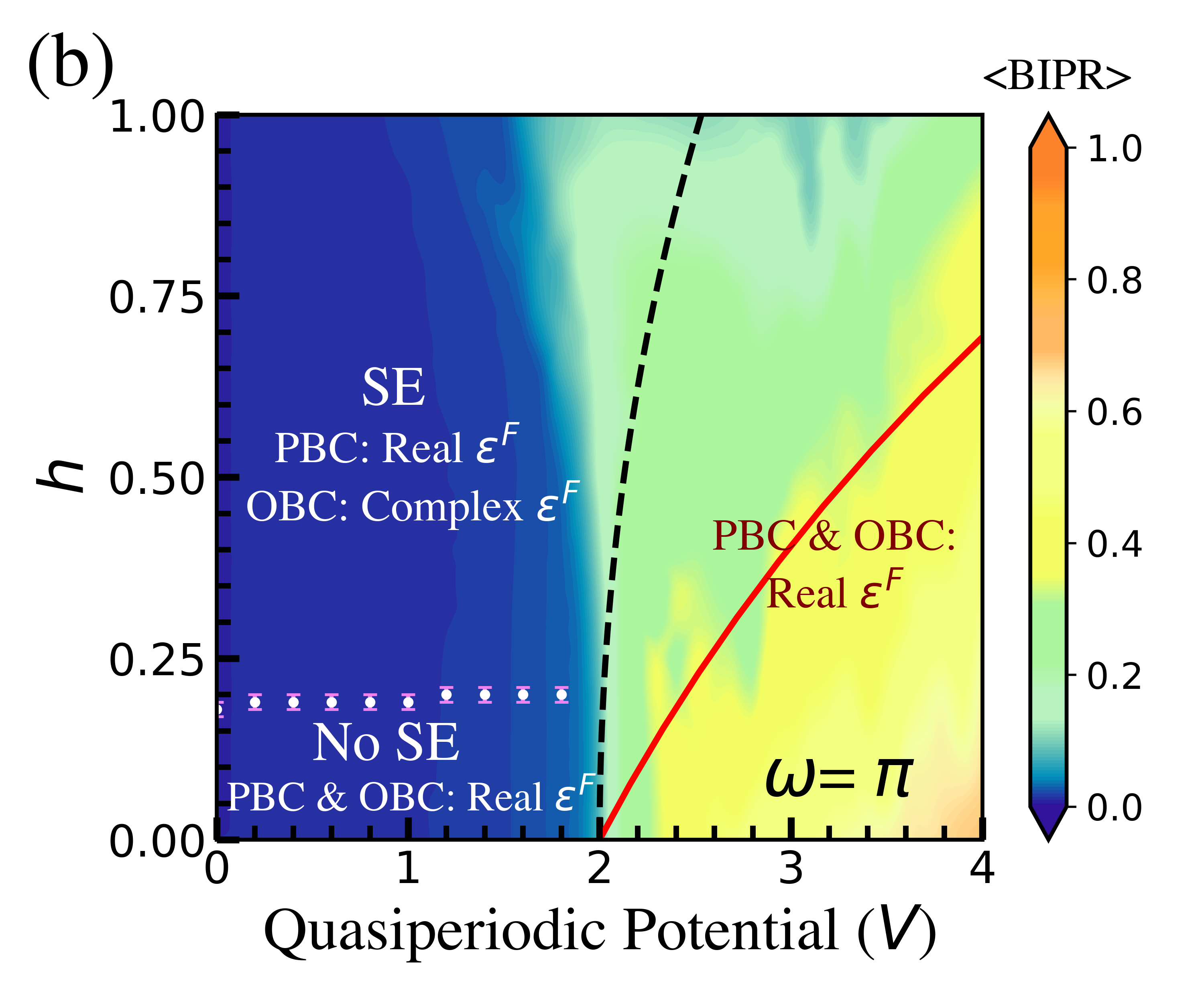}\\
	 		\includegraphics[width=0.244\textwidth,height=0.225\textwidth]{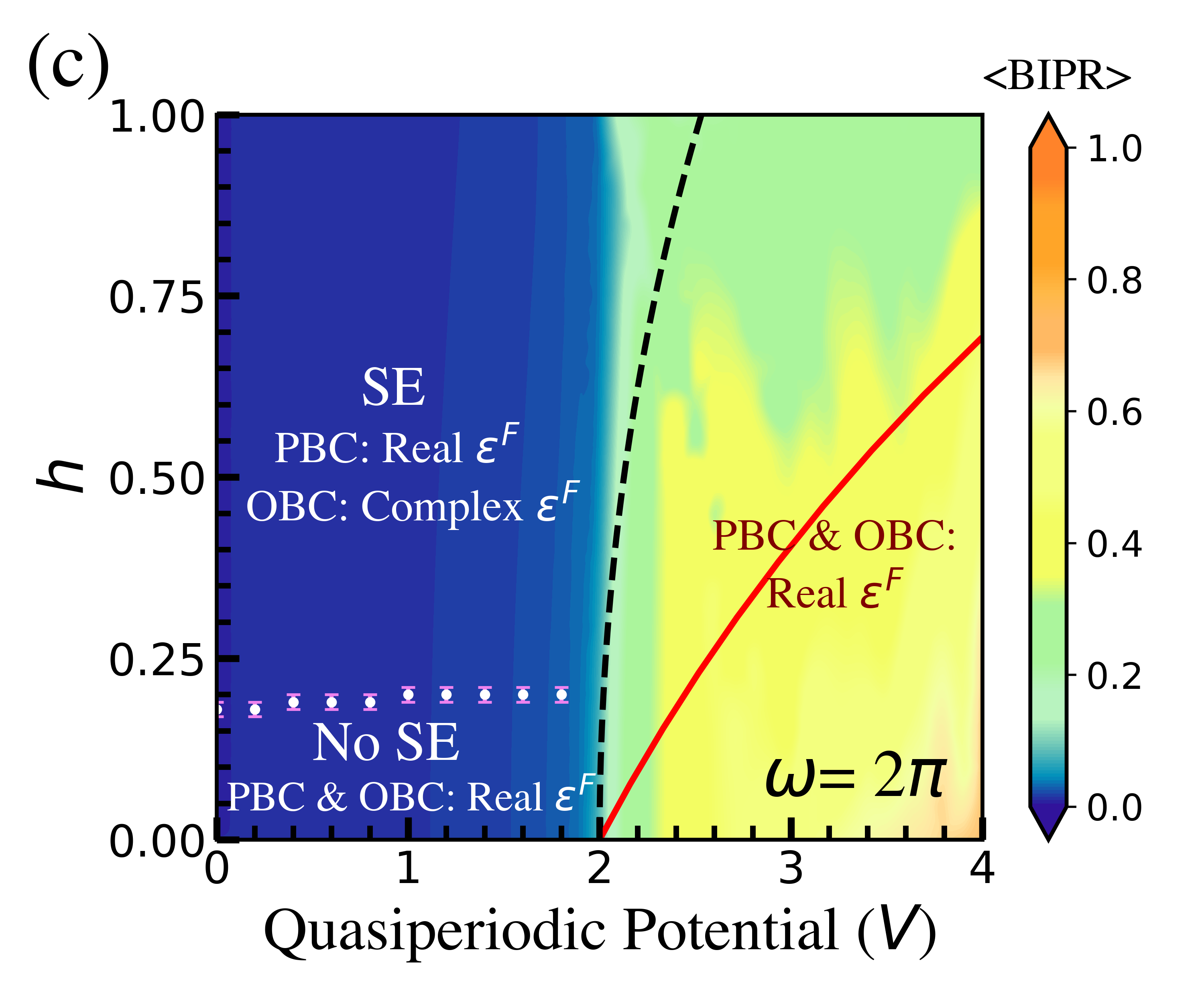}\hspace{-0.2cm}
	 		\includegraphics[width=0.244\textwidth,height=0.225\textwidth]{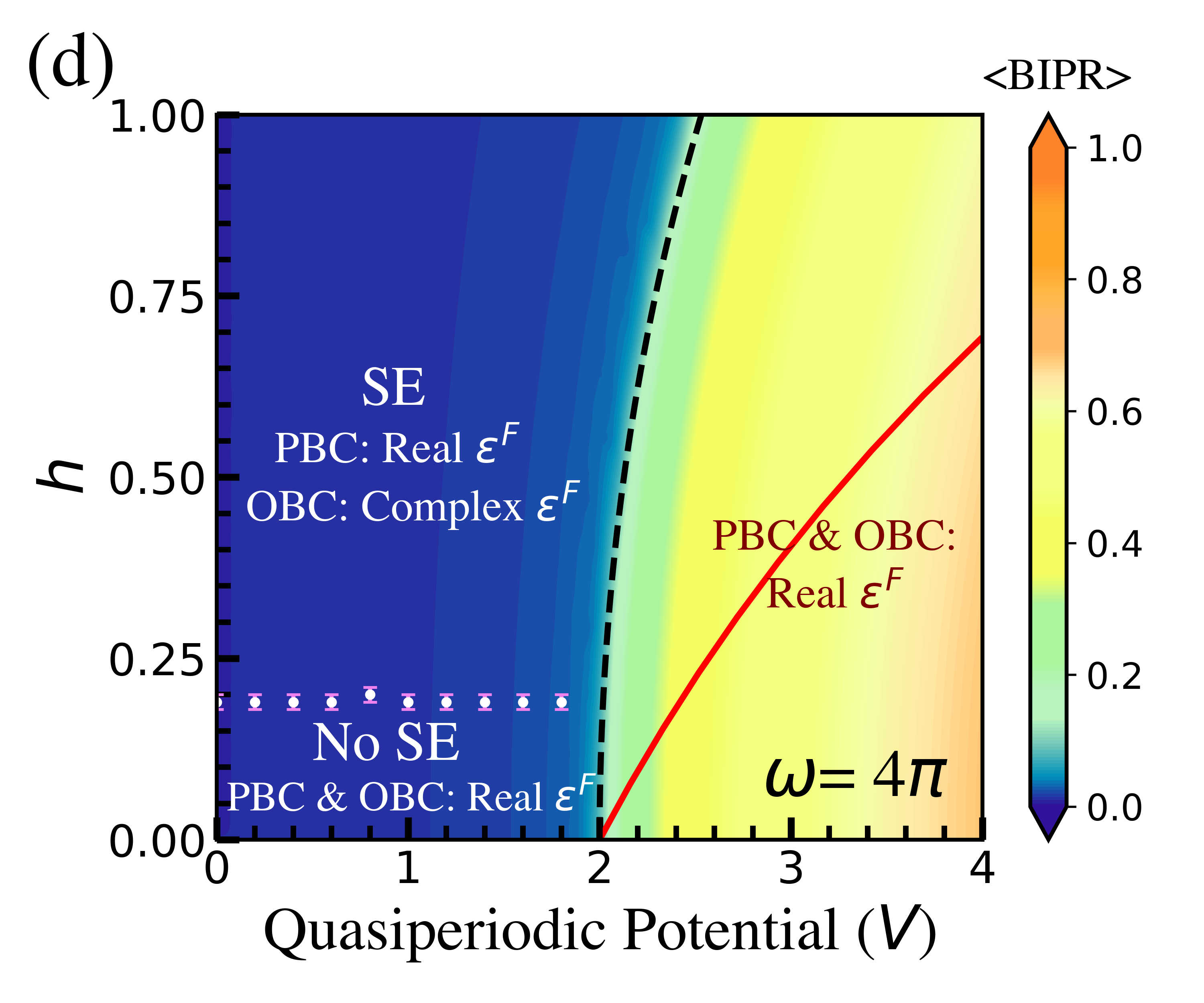}
	 		\caption{Projected $\left\langle \text{BIPR} \right\rangle$  as a function of $h$ and $V$ for the driven HN system at different driving frequencies: (a) $\omega=\pi/2$, (b) $\omega=\pi$, (c) $\omega=2\pi$, and (d) $\omega=4\pi$. The blue region in all the phase diagrams indicate the delocalized phase. The red solid line depicts the critical point for DL transition in the static HN Hamiltonian. The black dotted lines manifests the critical value under the drive as obtained analytically from the expression in Eq.~\ref{Eq:Critical_point}. The DL transition is demonstrated for a lattice with 144 sites and under the PBC. The white data markers superposed in all the phase diagrams separate the regions with and without the SE, obtained under OBC. The spectral behavior in the different regimes under both PBC and OBC are also indicated.}
	 		\label{Fig:Fig_1}
	 	\end{tabular}
	 \end{figure}
     \indent	 
	 The eigenvectors and eigenvalues of $U(T,0)$ on exact diagonalization yield the eigenspectrum of $\mathcal{H}_F$ given as,
	 \begin{eqnarray}
	 	U(T,0)=\displaystyle \sum_n E_n \ket{\psi_{nR}}\bra{\psi_{nL}}, \text{and} ~E_n=e^{-i\epsilon_n^FT} ,
	 	\label{Eq:Floquet_quasienergies}
	 \end{eqnarray}
 	 \noindent
	 where $E_n$, $\ket{\psi_{nR}}$ and $\ket{\psi_{nL}}$ are the eigenenergies and the right and left eigenvectors respectively. The Floquet quasienergies $\epsilon_n^F$ satisfy the relation $\mathcal{H}_F{\psi_n}=\epsilon_n^F{\psi_n}$ (defined modulo $\hslash\omega$). The solutions to the time dependent Hamiltonians to ascertain the physical properties can mostly be attained numerically using the Floquet eigenstates and quasienergies as described above.\\
	 \indent 
	 To identify the delocalized and localized phases in the system, the concept of Inverse Participation Ratio (IPR) \cite{Mirlin,Wessel} is widely used in the literature. The concept of IPR has been extended in non-Hermitian systems, where a new measure of bidirectional-IPR(BIPR) for an eigenstate labelled `$j$' has been introduced recently \cite{Wang_2019} and is defined as,
	 \begin{eqnarray}
	 	BIPR_j=\frac{\displaystyle\sum_{n=1}^{N} |\psi_{nL}^{j}\psi_{nR}^{j}|^2}{\Big(\displaystyle\sum_{n=1}^N |{\psi_{nL}^j\psi_{nR}^{j}}|\Big)^2}  ~~~~~,
	 	\label{Eq:BIPR}
	 \end{eqnarray}
	 where the sum of the weights of the wavevectors is over all the lattice sites indicated by $n$.
	 $\left\langle BIPR \right\rangle$ signifies the average of BIPR over all the eigenstates.
     The $\left\langle BIPR \right\rangle$ is $\mathcal{O}(L^{-1})$ when the states are completely delocalized, and  $\mathcal{O}(1)$ in the localized regime.
     For all the subsequent details and the findings on our periodically driven system, we have considered $J=1$ (in arb. units), $L=144$, and the Trotter time step as $\Delta t=0.001$, unless specifically stated.     
     \begin{figure}[]
     	\begin{tabular}{p{\linewidth}c}
     		\centering
     		\includegraphics[width=0.215\textwidth,height=0.215\textwidth]{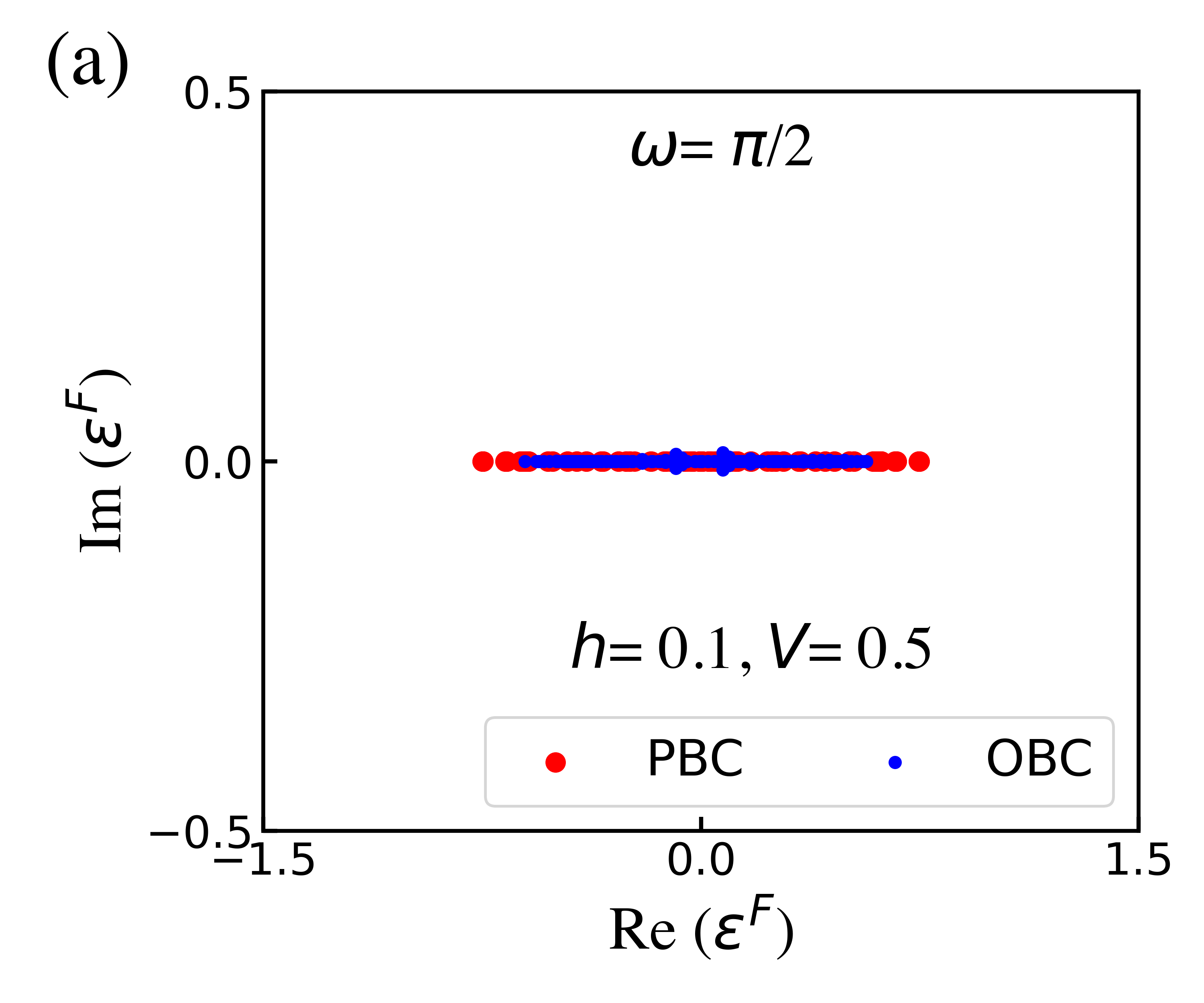}\hspace{-0.2cm}
     		\includegraphics[width=0.265\textwidth,height=0.215\textwidth]{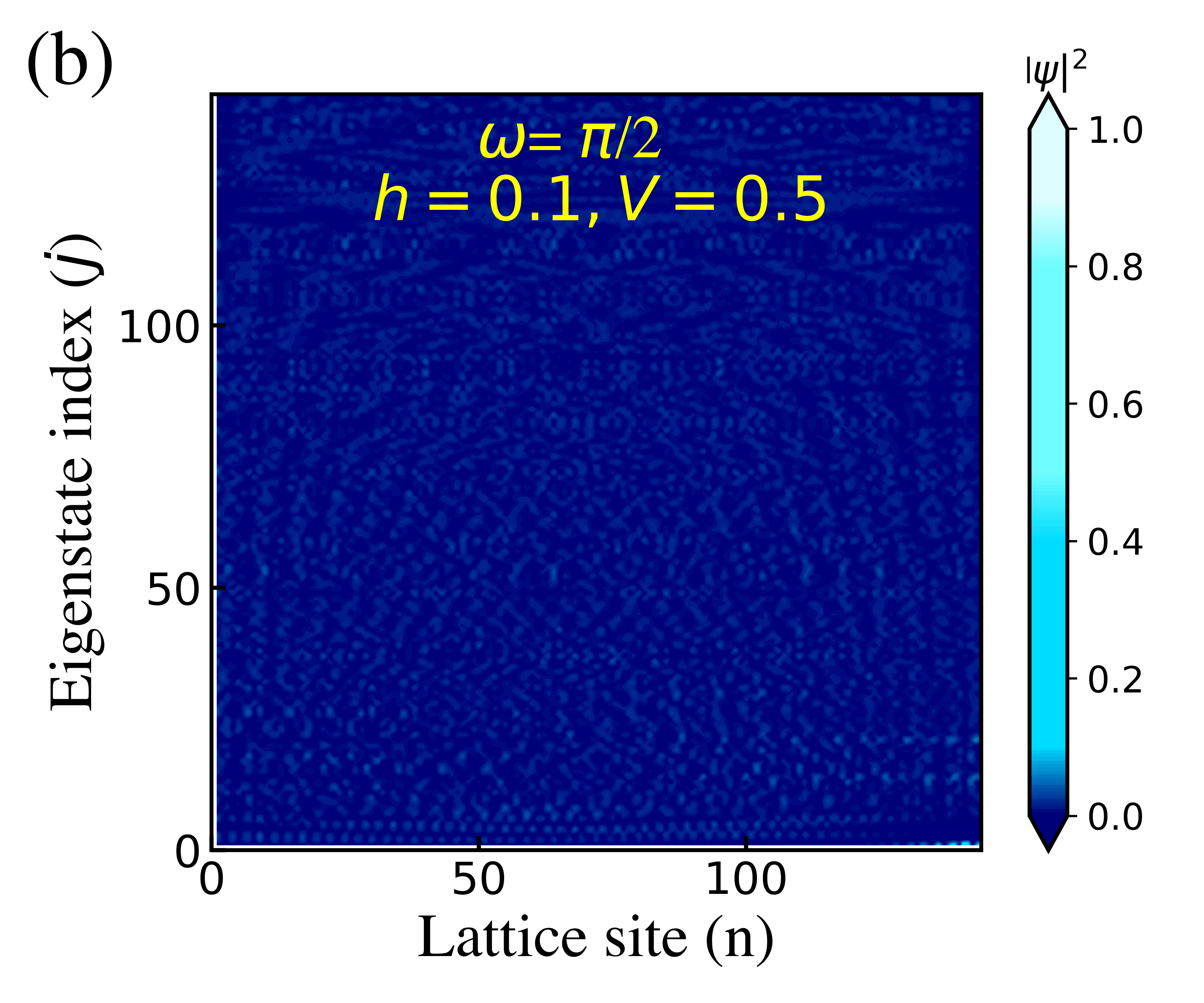}\\
     		\includegraphics[width=0.215\textwidth,height=0.215\textwidth]{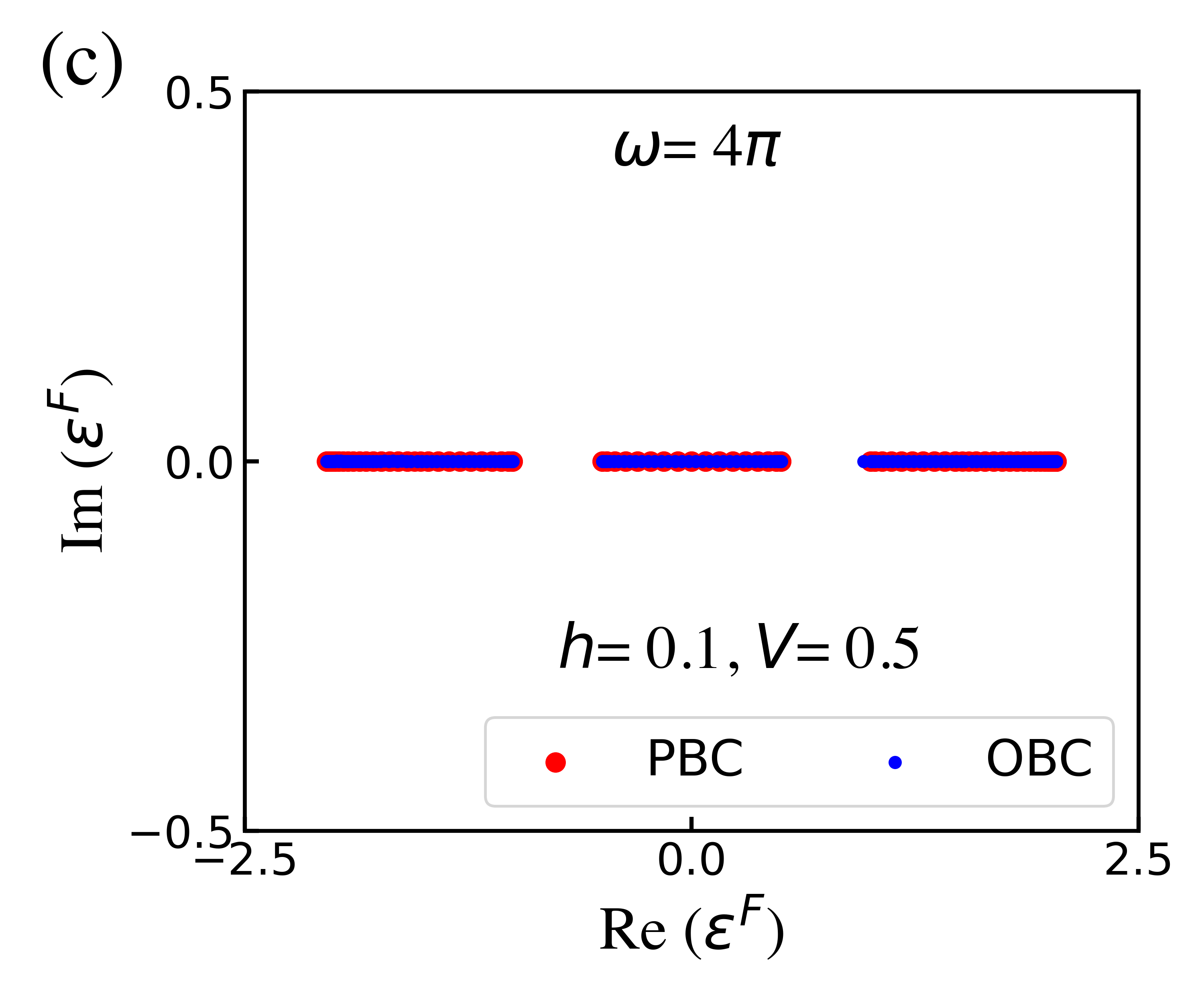}\hspace{-0.2cm}
     		\includegraphics[width=0.265\textwidth,height=0.215\textwidth]{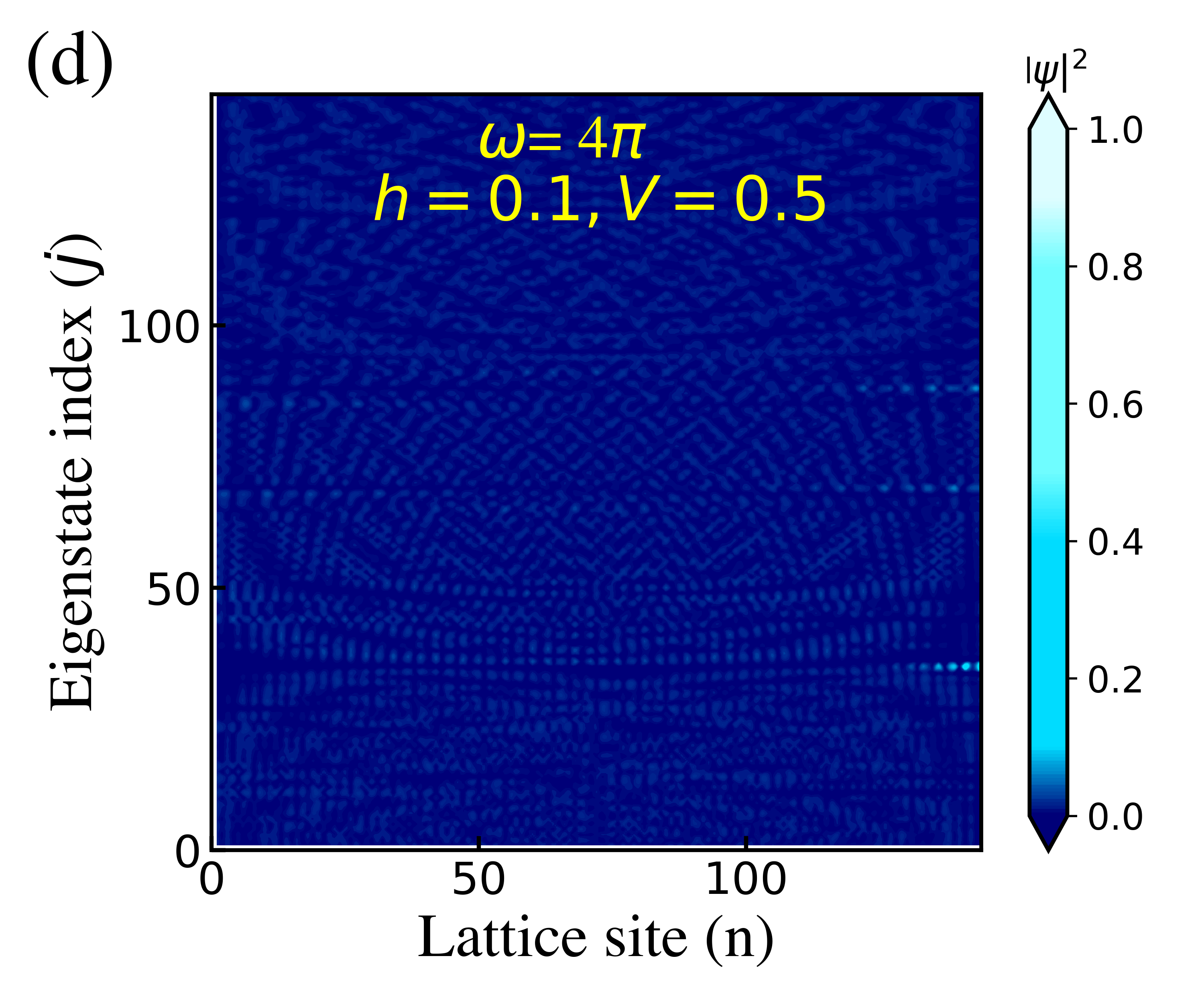}
     		\caption{The Floquet quasienergy spectrum under the PBC (in red) and OBC (in blue) for the driven system at a low value of the imaginary vector potential, i.e., $h=0.1$ and $V=0.5$ in (a) $\omega=\pi/2$ and (c) $\omega=4\pi$. (b,d) depict lattice-site resolved $|\psi|^2$ for all the eigenstates corresponding to the two parameters in (a) and (c) respectively, indicating the absence of SE under the OBC. $L=144$ in all the cases.}
     		\vspace{-0.9cm}
     		\label{Fig:Fig_2}
     	\end{tabular}
     \end{figure}
 
 	\begin{figure}[]
     	\begin{tabular}{p{\linewidth}c}
     		\centering
     		\includegraphics[width=0.215\textwidth,height=0.215\textwidth]{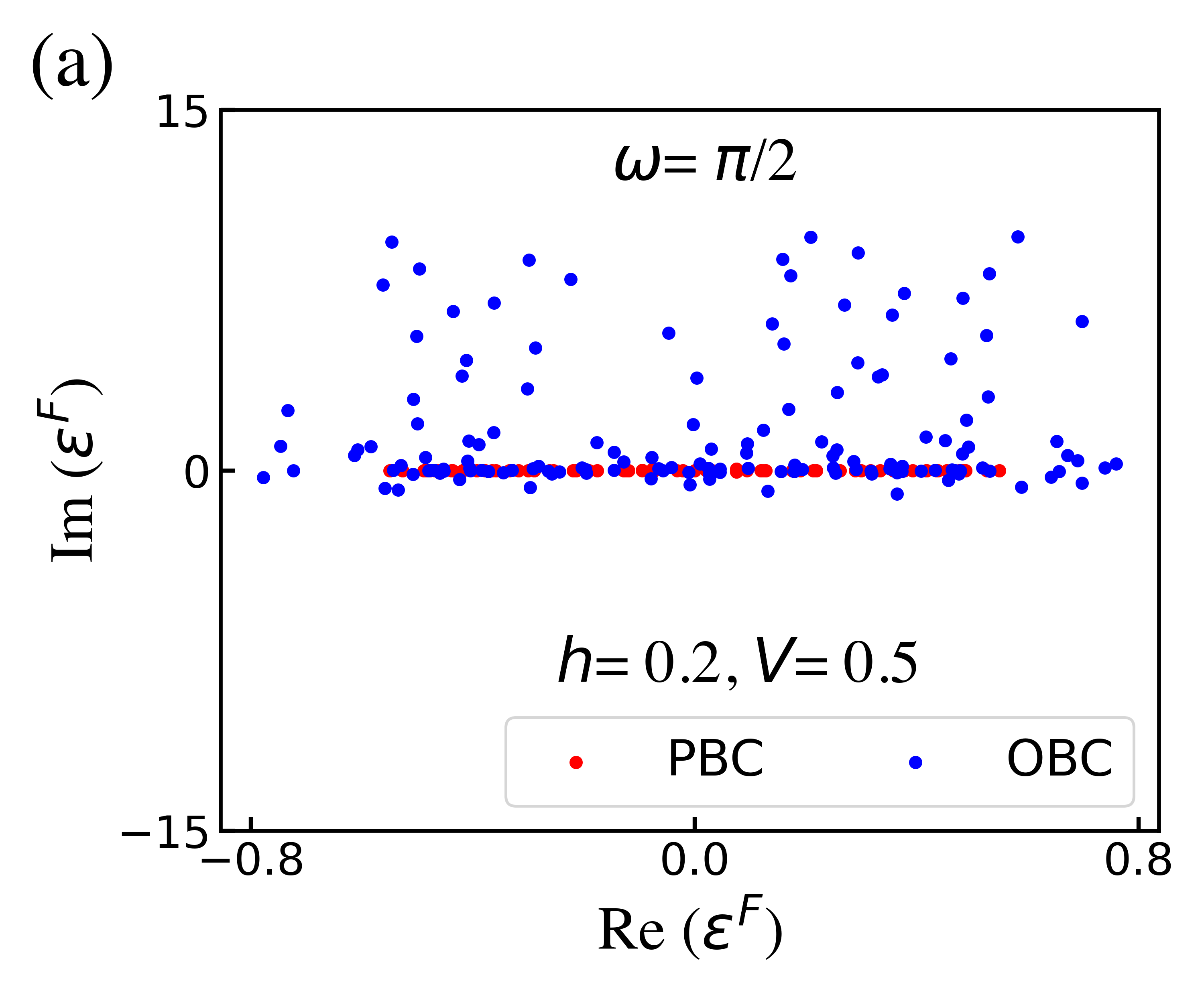}\hspace{-0.2cm}
     		\includegraphics[width=0.265\textwidth,height=0.215\textwidth]{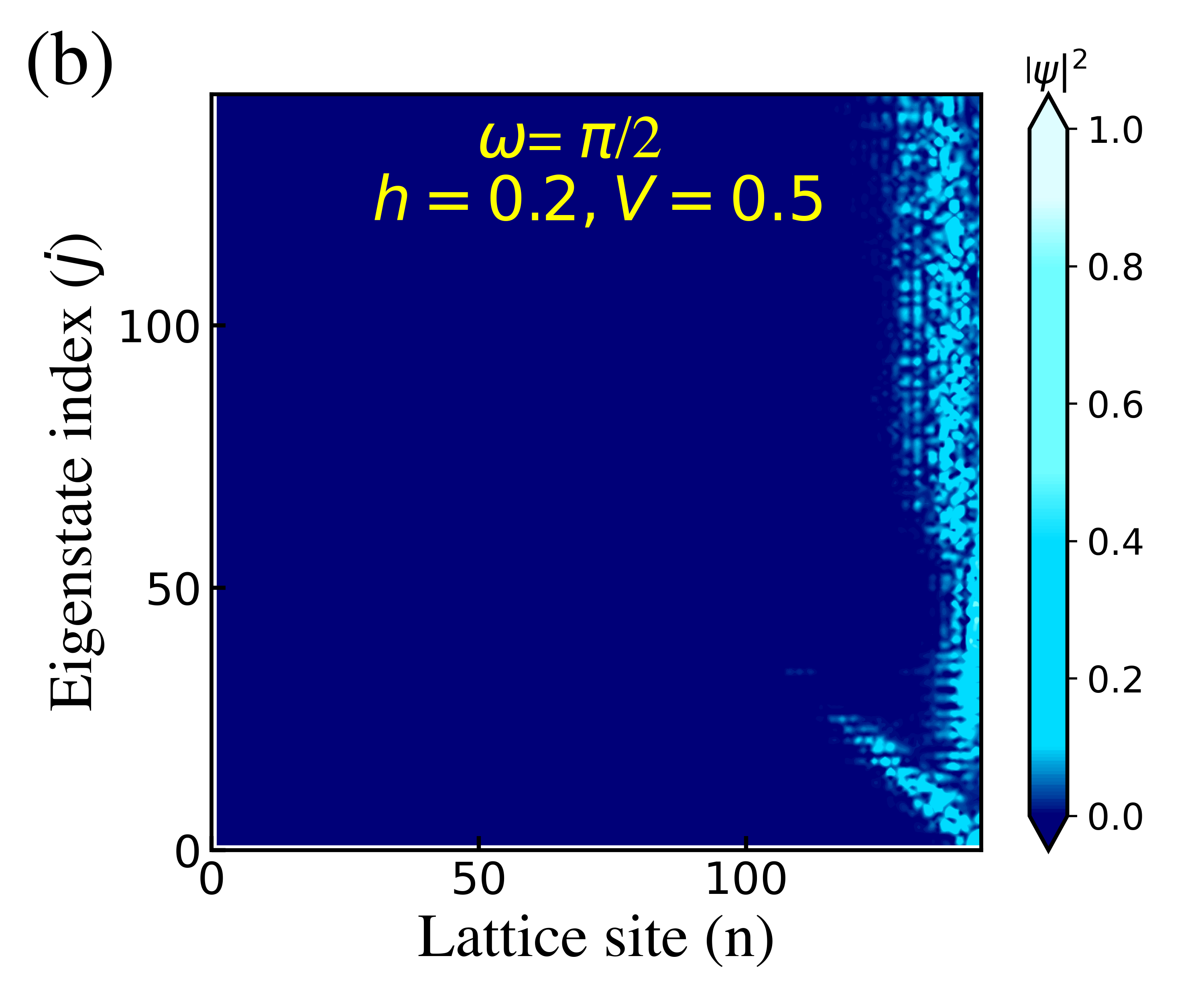}\\
     		\includegraphics[width=0.215\textwidth,height=0.215\textwidth]{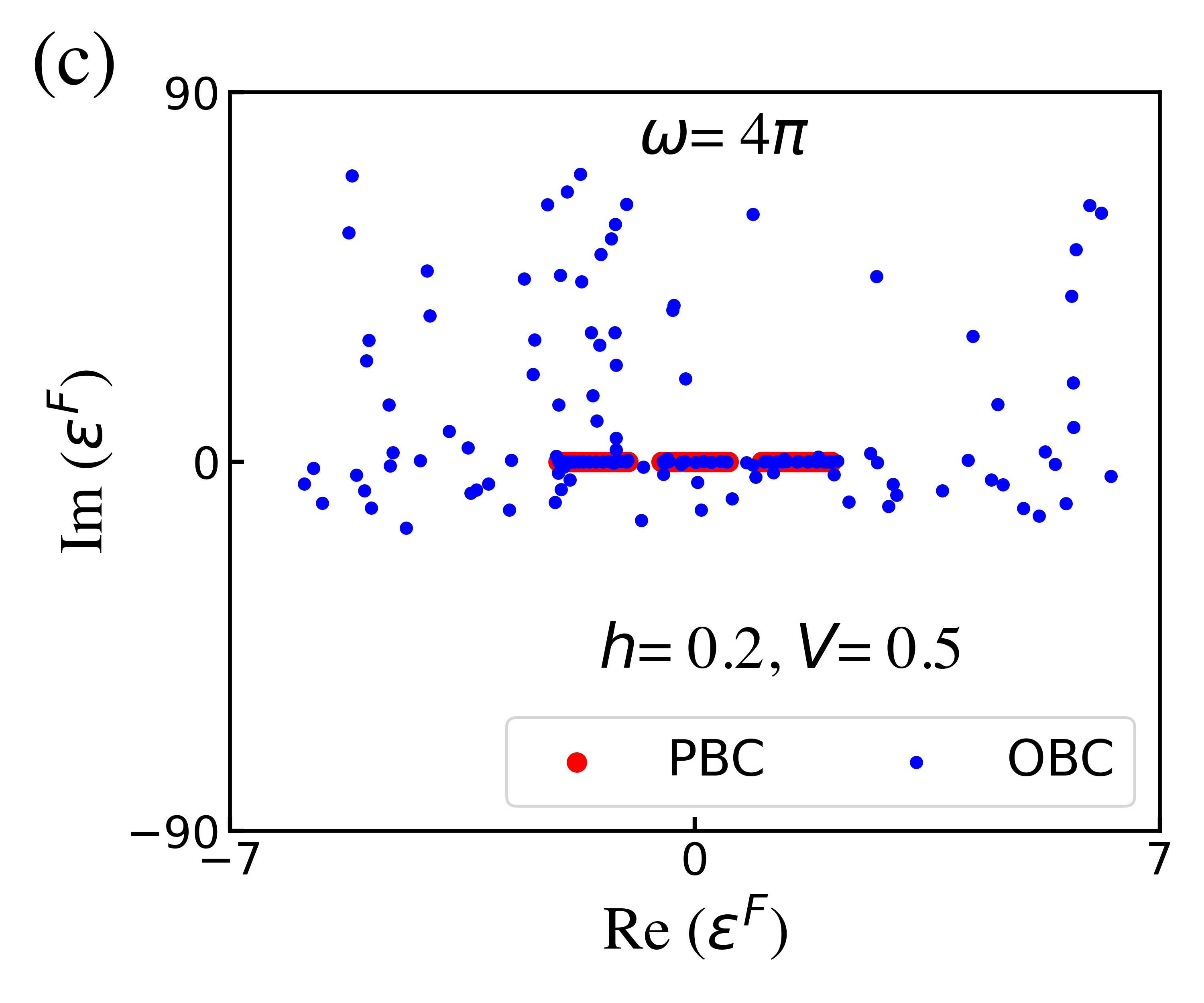}\hspace{-0.2cm}
     		\includegraphics[width=0.265\textwidth,height=0.215\textwidth]{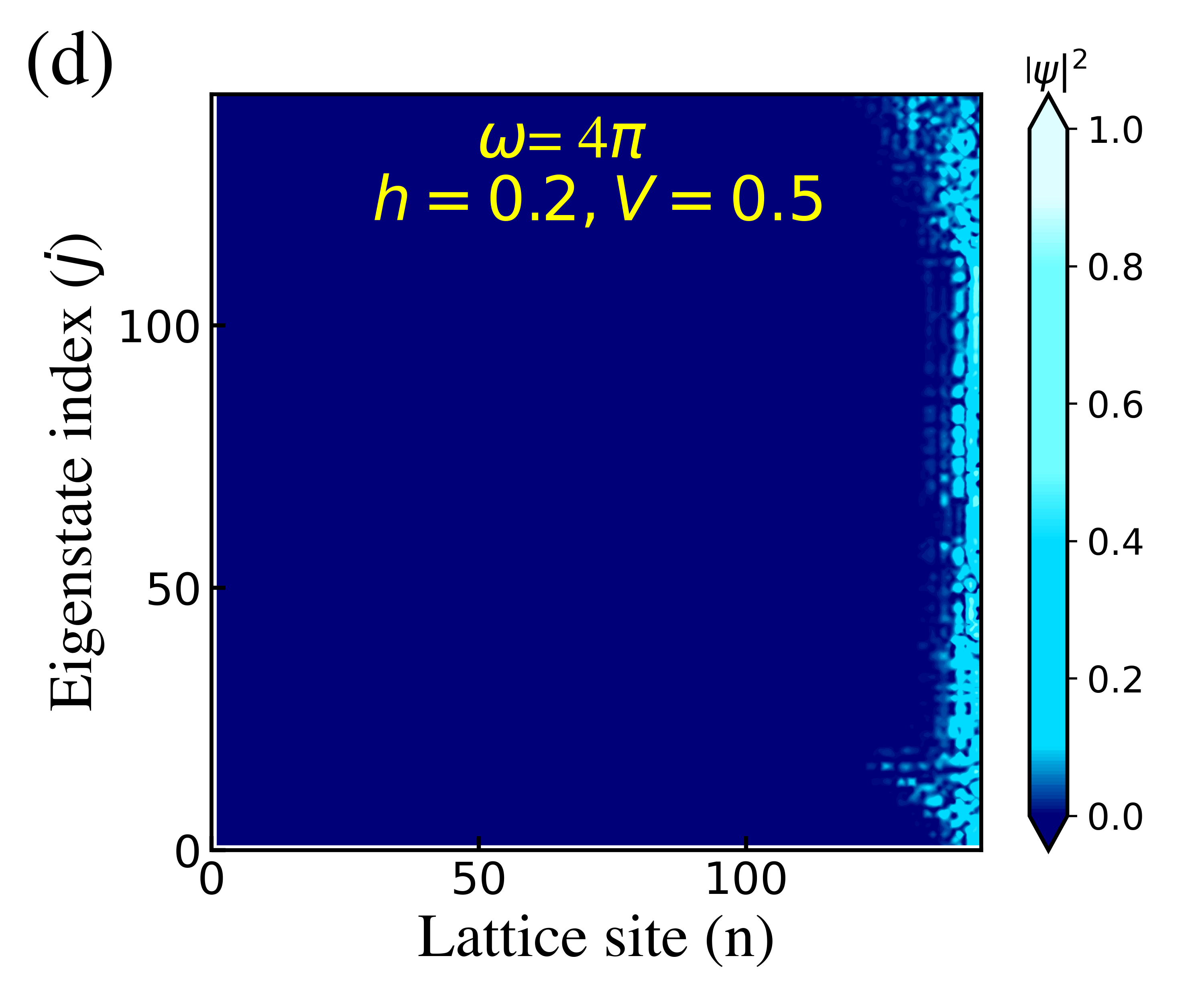}
     		\caption{The Floquet quasienergy spectrum under the PBC (in red) and OBC (in blue) for the driven system at $h=0.2$ and $V=0.5$ in (a) $\omega=\pi/2$ and (c) $\omega=4\pi$. (b,d) demonstrates the lattice-site resolved $|\psi|^2$ for different eigenstates for the two parameters corresponding to (a) and (c), demonstrating the presence of SE under the OBC. The lattice consists of $144$ sites in all the cases.}
     		\vspace{-0.9cm}
     		\label{Fig:Fig_3}
     	\end{tabular}
     \end{figure}
     \indent \textit{Results and discussions-} 
     The aim of this work is achieved in two steps.
     We first explore the behavior of DL transition of the time-dependent HN system under PBC in the parameter space 
     of $h$ and $V$ for different driving frequencies. In the subsequent analysis, we assess the sensitivity of the
     energy spectrum to the boundary conditions alongwith the investigation of the existence of SE. \\
     \indent
     Figs.~\ref{Fig:Fig_1}(a-d) illustrate the phase diagram of the time-periodic version of the HN Hamiltonian as discussed in Eq.~\ref{Eq:Hamiltonian} under the PBC. In sharp contrast to the static system, interestingly, we
     observe from  Figs.~\ref{Fig:Fig_1}(a-d) that the DL transition (indicated by a transition from blue to a different color) does not follow the critical value of the undriven system (indicated by the solid red line) determined from the condition $V_c=\text{max}[Je^{h},Je^{-h}]$, as demonstrated in Ref.~\cite{Jiang}.
     From the phase diagrams, it is evident that the transition from delocalized to localized states strongly depends
     on the driving frequency. \\
     \indent
     In the following, we demonstrate that the critical point of the DL transition under the PBC can be obtained analytically in the regime of a large frequency of the drive. The analytical expression of the phase boundary 
     is derived by retrieving an effective Floquet Hamiltonian upon time-averaging the original Hamiltonian as discussed in earlier works \cite{Goldman,Bukov,Eckardt}. The effective Hamiltonian, which can only be obtained 
     when the frequency of the drive is large enough, is written as follows,
     \begin{eqnarray}
     	\mathcal{H}_{F}=\frac{1}{T}\int_{0}^{T}\mathcal{H}(t) dt.
     	\label{Eq:High_frequency_Hamiltonian}
     \end{eqnarray}	
	  To evaluate $\mathcal{H}_{F}$ as given in Eq.~\ref{Eq:High_frequency_Hamiltonian}, we first consider the hopping towards the right, which is given by,
     \begin{eqnarray}
     	\frac{1}{T}\int_{0}^{T} \displaystyle\sum_{n} (J e^{h\text{cos}(\omega t)} c^\dag_{n+1} c_{n}) dt.
     	\label{Eq:HF_part1}
     \end{eqnarray}	
     In the above equation $\int_{0}^{T} e^{h\text{cos}(\omega t)} dt$ can be identified as the generating function
     of the modified Bessel equation, i.e., $\int e^{\frac{z}{2}(x+\frac{1}{x})} dx$ with $z=h$ and $x=\text{exp}(i \omega t)$. Integration over the entire time period yields the effective Floquet Hamiltonian for this part, and is given as,
     \begin{eqnarray}
     	\displaystyle\sum_{n} \Big(J\mathcal{I}_0(h)\Big) c^\dag_{n+1} c_{n},
     	\label{Eq:HF_part1_exp}
     \end{eqnarray}
     where $\mathcal{I}_0$ is the zeroth order modified Bessel function. It is interesting to note that the above expression is independent of $\omega$. Similarly, for the hopping in the opposite direction (towards the left), the other part of the effective Hamiltonian can be written as,
     \begin{eqnarray}
     	\displaystyle\sum_{n} \Big(J\mathcal{I}_0(-h)\Big) c^\dag_{n} c_{n+1}.
     	\label{Eq:HF_part2_exp}
     \end{eqnarray}
     It is easy to see that $\mathcal{I}_0(h)=\mathcal{I}_0(-h)$. The effective Floquet Hamiltonian can then be explicitly written as,
     \begin{eqnarray}
     	\mathcal{H}_{F}= \displaystyle\sum_{n} \Big(J\mathcal{I}_0(h)\Big) (c^\dag_{n+1} c_{n}
     	+c^\dag_{n} c_{n+1})+\nonumber \\
     	\sum_{n} V \text{cos} (2\pi\alpha n) c^\dag_{n} c_{n}.~~~~~~~~~~~~~~
     	\label{Eq:Eff_Ham}
     \end{eqnarray}
     \indent Remarkably, from Eq.~\ref{Eq:Eff_Ham}, it becomes quite evident that the effective Floquet
     Hamiltonian for the HN systems with a drive in the magnetic vector potential reduces to a Hermitian AAH Hamiltonian with a rescaled hopping amplitude given as $J\mathcal{I}_0(h)$ in the regime of high frequency.
     Thus, we expect the DL transition to occur at a critical value of the quasiperiodic potential determined by the self-duality of the AAH Hamiltonian in the Hermitian limit, expressed as,
     \begin{eqnarray}
     	V_c=2J', ~~\text{where}~~ J'=J\mathcal{I}_0(h).
     	\label{Eq:Critical_point}
     \end{eqnarray}
 	 In Figs.~\ref{Fig:Fig_1}(a-d), we have indicated the analytically obtained phase boundary in Eq.~\ref{Eq:Critical_point} by a black dotted line. It is clear that as we approach to a higher 
 	 value of the driving frequency ($\omega=4\pi$), the numerically determined critical value for the DL transition
 	 agrees excellently with the analytical result in the entire parameter space. However, for lower driving frequencies, the numerical critical points follow the analytical expression only upto a certain strength of the asymmetry. 
     In addition, the effect of the periodic drive in such systems is to shift the DL transition to a lower value of $V_c$, as compared to its static counterpart (demonstrated by the red line in Figs.~\ref{Fig:Fig_1}(a-d)). Moreover, our analytical result suggests that the Floquet quasienergies in the entire parameter space of $h$ and $V$ should be real for high driving frequencies under PBC, due to the exact mapping of the original non-Hermitian system to the Hermitian AAH Hamiltonian. We have numerically verified this assertion. Suprisingly, however, the reality of the quasienergies persists for all frequencies of the drive.
     This is in complete contrast to the static HN Hamiltonian under the PBC, where the DL transition is concurrent 
     with a spectral transition from complex to real. In our case, the existence of the real quasienergies  irrespective of the strength of the potential (refer to Fig.~\ref{Fig:Fig_S1} of the supplemental material for more details) and non-Hermiticity, which is the hallmark for EU, suggests that our system becomes dynamically stable after a stroboscopic driving period.\\
     \indent 
     In non-Hermitian systems a change in the boundary condition from PBC to OBC drastically alters the behaviour of
     electronic states and eigenenergies, especially in the delocalized phase. In the static HN counterparts, it has been well demonstrated that under the OBC, in the delocalized regime, the eigenenergies lie on the real axis \cite{Sato}, alongwith the appearance of SE. Naturally, one of the most important question that arises is 
     whether such a correspondence between the quasienergies and SE exist under the time-periodic drive. We address this question in the subsequent discussion.      
     \begin{figure}[]
     	\begin{tabular}{p{\linewidth}c}
     		\centering
     		\includegraphics[width=0.30\textwidth,height=0.265\textwidth]{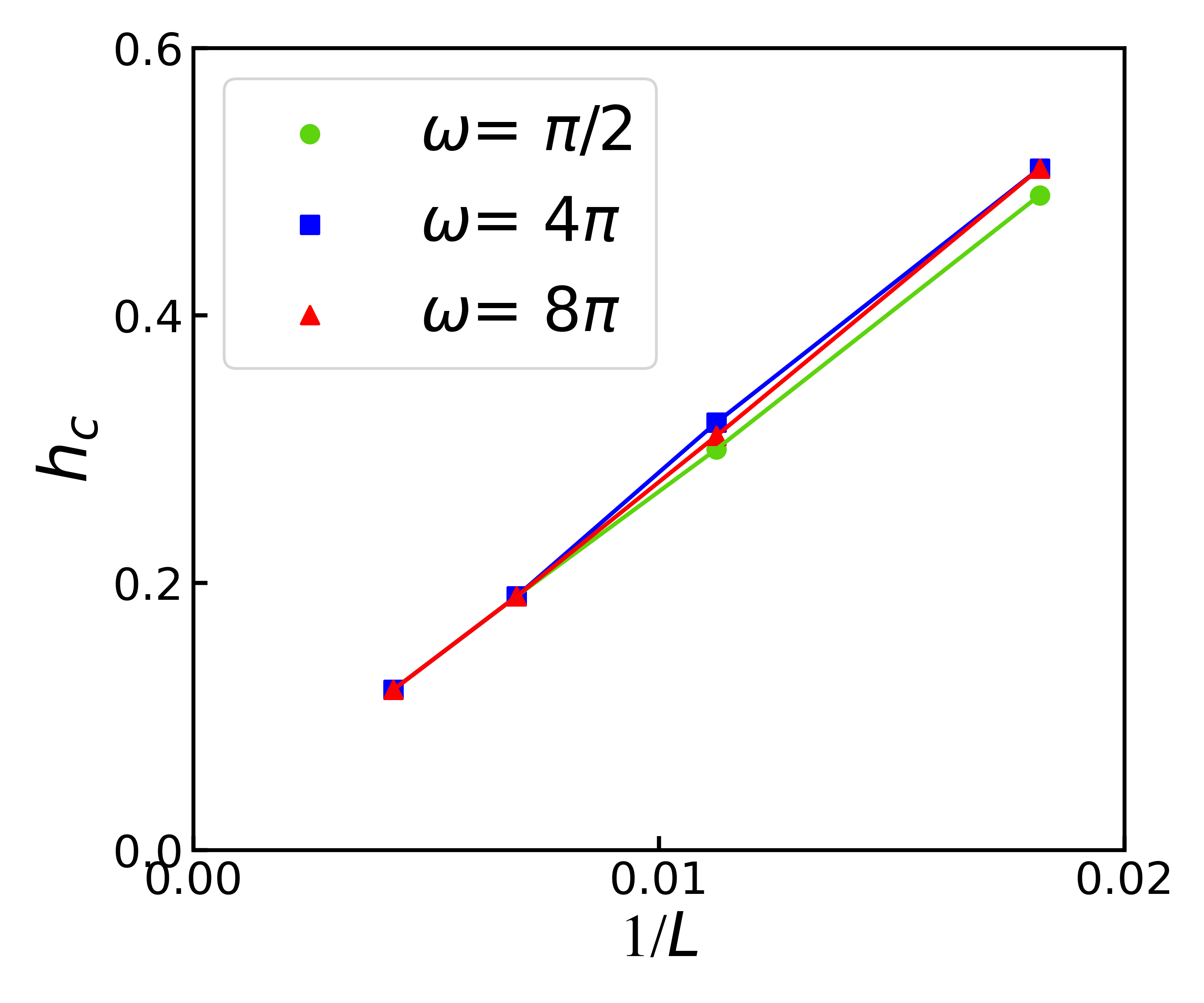}
     		\vspace{-0.4cm}
     		\caption{System-size behavior of $h_c$ for the onset of SE as a function of inverse lattice size in different regimes of the Floquet drive, i.e., at $\omega=\pi/2$ (green), $\omega=4\pi$ (blue) and $\omega=8\pi$ (red). The data are obtained for $L$= 55, 89, 144 and 233 under the OBC.}
     		\vspace{-0.9cm}
     		\label{Fig:Fig_4}
     	\end{tabular}
     \end{figure}
     \indent
     In analogy with the static system, one can naively anticipate that the SE would persist in the entire delocalized regime (represented in blue in Figs~\ref{Fig:Fig_1}(a-d)) retaining the reality of the Floquet quasienergies. However, we demonstrate that this expectation is in complete contradiction when the system 
     is driven.
     From Figs.~\ref{Fig:Fig_2} (a,c) it is evident that for weak asymmetry in the hopping, the periodic drive leads to real Floquet quasienergies under both the periodic and open boundary conditions. However, remarkably, we find that under the OBC, the time-periodic drive demolishes the SE regardless of the driving frequency (Figs.~\ref{Fig:Fig_2} (b,d)), contrary to the undriven systems where the SE exists for an arbitrarily weak  strength of the imaginary magnetic vector potential.\\
     \indent To further understand the dependence of the SE on $h$, we consider a slightly greater amplitude of the asymmetricity in the Hamiltonian. In this case, the quasienergy spectrum under the PBC still exhibits real eigenvalues, similar to a Hermitian system, as demonstrated in Figs.~\ref{Fig:Fig_3} (a,c), manifesting EU as indicated previously. This is in accordance with the finding in Eq.~\ref{Eq:Eff_Ham}. However, with an increase in the value in $h$, the spectrum under OBC lies on the complex plane and does not satisfy the EU condition. Surprisingly, when the spectrum becomes complex under OBC, the system exhibits SE irrespective of the driving frequency, as illustrated in Figs.~\ref{Fig:Fig_3} (b,d). This is in stark contrast to the static systems. From these findings, it is clear that there exists a critical value $h_c$ at which the SE appears. We have 
     numerically determined $h_c$ in the entire parameter space of $h$ and $V$. Our findings are summarized in 
     Figs.~\ref{Fig:Fig_1}(a-d) for different frequencies of the drive. It is remarkable that $h_c$ is independent of the driving frequency and the quasiperiodic potential. The numerical value of $h_c$ is estimated to be 
     $h\simeq0.20$ with an error of $\pm 0.01$. The preceding discussions clearly indicate that the emergence of EU and disappearance of SE are closely linked in such driven systems. The determination of $h_c$ for the different driving frequencies has been elucidated in Fig.~\ref{Fig:Fig_S2} of the supplemental material.
     \\
     \indent
     It is important to note that the critical strength of the asymmetricity for the onset of SE in such driven systems depends strongly on the size of the lattice. In order to get a comprehensive idea on this system size
     dependence, in Fig.~\ref{Fig:Fig_4} we present $h_c$ as a function of $1/L$ for different driving frequencies. 
     It is evident that $h_c$ scales inversely with the system size. In addition, $h_c$ becomes independent of the driving frequency when the size of the lattice is large enough. Furthermore, it can be easily observed that $h_c$ approaches to the static limit of HN Hamiltonian for $L\to\infty$.     
    \begin{figure}[]
    	\begin{tabular}{p{\linewidth}c}
    		\centering
    		\includegraphics[width=0.30\textwidth,height=0.30\textwidth]{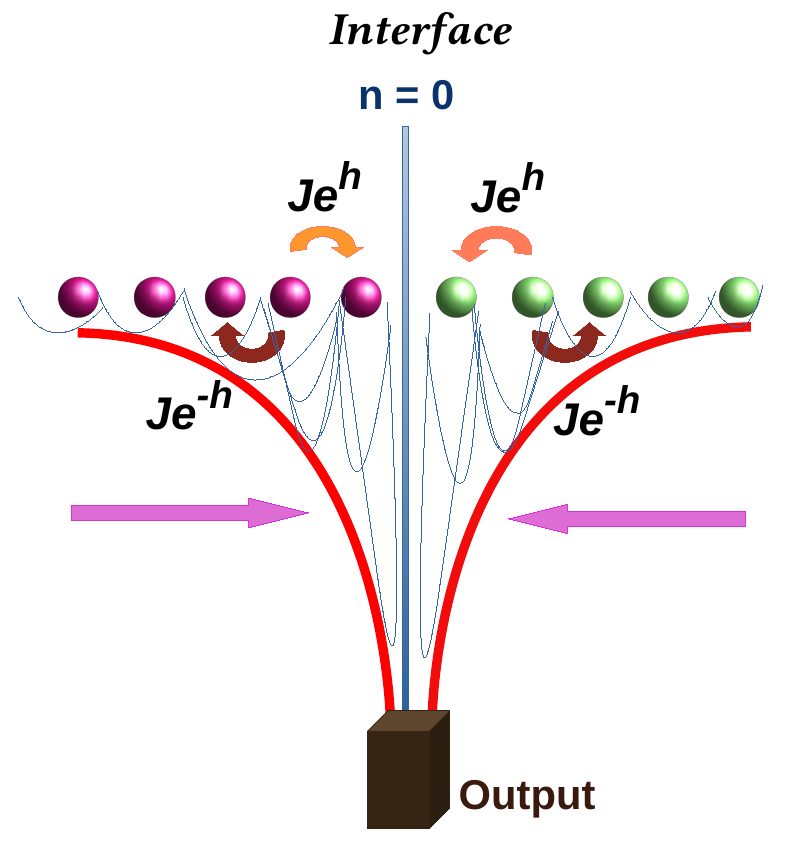}
    		\vspace{-0.4cm}
    		\caption{ Schematic of light funneling using two oppositely directed static HN chains. The interface is formed at $n=0$.
    		The chains on the left and right side of the interface have greater unidirectionality towards the right and left directions respectively.
    		The output light traverses through the funnel-like structure where it is collected.}
    		\vspace{-0.9cm}
    		\label{Fig:Fig_5}
    	\end{tabular}
    \end{figure}
    
    \indent \textit{Switch for light funneling-}
    In this discussion, we propose a possible experimental implementation of the system considered in this work 
    from the point view of its application in controlling the light funneling effect.
    Similar to the recent realizations of the non-reciprocal lattices in photonic systems as demonstrated in Refs.~\cite{Longhi_SE,Weidemann}, we consider a set-up with two optical fibers with anisotropic hopping in either directions, seperated by an interface as illustrated in Fig.~\ref{Fig:Fig_5}.
    The driven HN system can be replicated by changing the anisotropy in the hopping with the help of a frequency dependent beam-splitter \cite{Makarov}, and selectively using the desired beam of light.
    Such an unidirectionality in the two fibers causes the incident light that is impinged on the lattice to be pushed towards the interface due to SE, resulting in a funneling effect of the output light. Since the value of $h$ can be tuned with the anisotropic beam splitter and our results suggest the existence of $h_c$ for the onset of SE, it is easy to understand that the SE can be
    turned on or off by controlling the beam splitter, effectively acting as a switch to tune the light-funneling 
    effect.\\
    \indent \textit{Conclusions-} In conclusions, this Letter demonstrates that a time-periodic drive can 
    fundamentally alter the behaviour of the paradigmatic quasiperiodic HN systems. By introducing a drive, we 
    unfold that under PBC, the Floquet Hamiltonian in the large frequency limit becomes equivalent to a Hermitian AAH counterpart, giving rise to the \textit{extended unitarity} and real Floquet quasienergies in the entire paramter space of the Hamiltonian, in stark contrast to the static limit. We find that the driven system undergoes a DL transition similar to the static case, albeit at a lower strength of the potential for a given asymmtetry in the hopping, determined by the self-duality condition of the effective Hermitian AAH Hamiltonian. Under OBC, however,
    we find that the EU condition survives only up to a critical value of the asymmetry in the hopping, $h_c$, along with the disappearance of SE within that asymmetry. In complete contrast to the static limit, the SE appears above this critical value with complex Floquet quasienergies. Remarkably, for a given system size, $h_c$ is found to be independent of the driving frequency and quasiperiodic potential. Furthermore, we find that $h_c$ scales inversely with the system size approaching to the static behaviour in the thermodynamic limit. Our work deepens the understanding of the 
    DL transition under the PBC, Floquet quasienergies and its connection to the SE under the OBC in HN systems. 
    Finally, we provide an experimental set-up that can exploit the findings of this Letter to control the 
    light-funelling mechanism. \\
     \indent
     A.C. acknowledges the financial support received from the Council of Scientific $\&$ Industrial Research (CSIR)-HRDG, India,
     via File No.- 09/983(0047)/2020-EMR-I.
     The data computation in this work were carried out in the cluster procured from SERB (DST), India (Grant No. EMR/2015/001227) and
     using the High Performance Computing (HPC) facilities of the National Institute of Technology Rourkela.
     
    \bibliography{reference.bib}
%%%%%%%%%%%%%%%%%%%%%%%%%%%%%%%%%%%%%%%%%%%%%%%%%%%%%%%%%%%%%%%%%%%%%%%%%%%%%%%%%%%%%%%%%%%%%%%%%%%%%%%%%%%%%%%%
%\title{\textbf{Supplemental material for ``Extended unitarity and absence of skin effect in periodically driven systems''}}
%\author{Aditi Chakrabarty}
%\email{aditichakrabarty030@gmail.com}
%\affiliation{Department of Physics and Astronomy, National Institute of Technology, Rourkela, Odisha-769008, India}
%\author{Sanjoy Datta}
%\email{dattas@nitrkl.ac.in}
%\affiliation{Department of Physics and Astronomy, National Institute of Technology, Rourkela, Odisha-769008, India}
%\date{\today}
%\maketitle
\newpage
\begin{center}
Supplemental material for ``Extended unitarity and absence of skin effect in periodically driven systems''	
\end{center}

\noindent\textbf{S.1. Existence of extended unitarity in the localized regime}\\

\begin{figure*}[]
	\begin{tabular}{p{\linewidth}c}
		\centering
		\includegraphics[width=0.244\textwidth,height=0.225\textwidth]{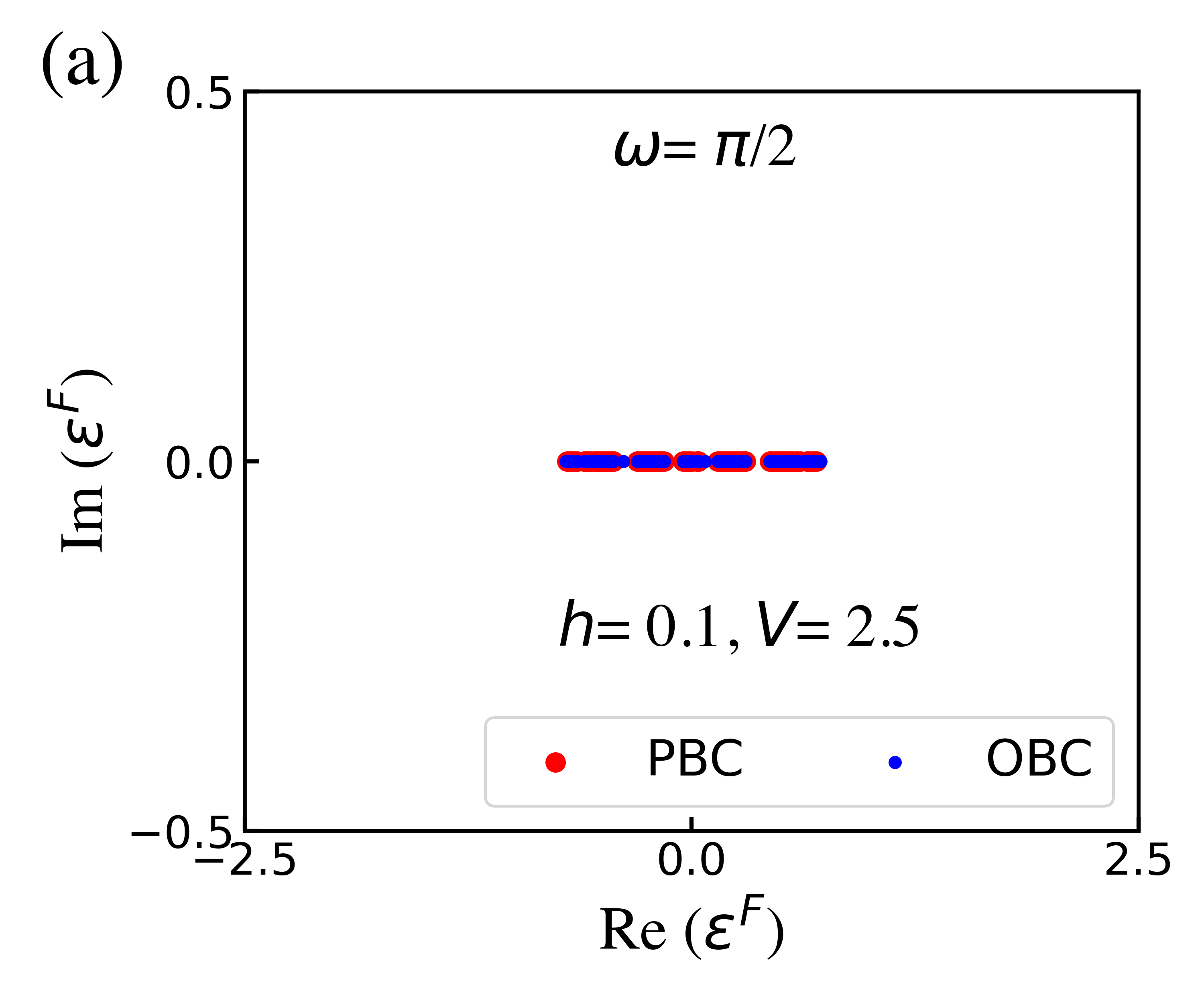}\hspace{-0.2cm}
		\includegraphics[width=0.244\textwidth,height=0.225\textwidth]{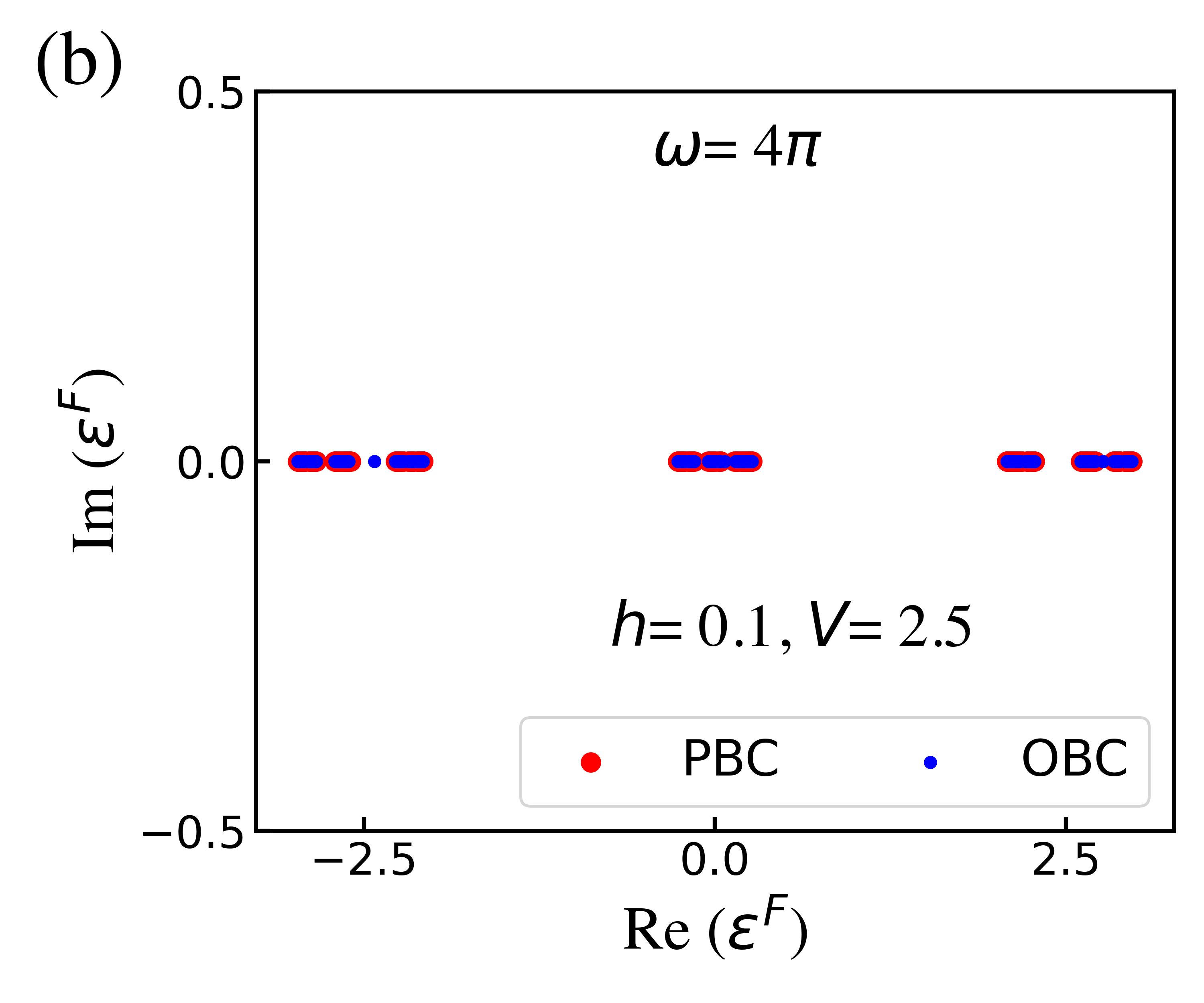}
		\caption{The Floquet quasienergy spectrum in the complex plane under the PBC (in red) and OBC (in blue) at $h=0.1$ and $V=2.5$ (localized regime) at: (a) $\omega=\pi/2$, (b) $\omega=4\pi$. $L=144$ as considered in the main text.}
		\label{Fig:Fig_S1}
	\end{tabular}
\end{figure*}

\begin{figure*}[b]
	\centering
	\includegraphics[width=0.247\textwidth,height=0.225\textwidth]{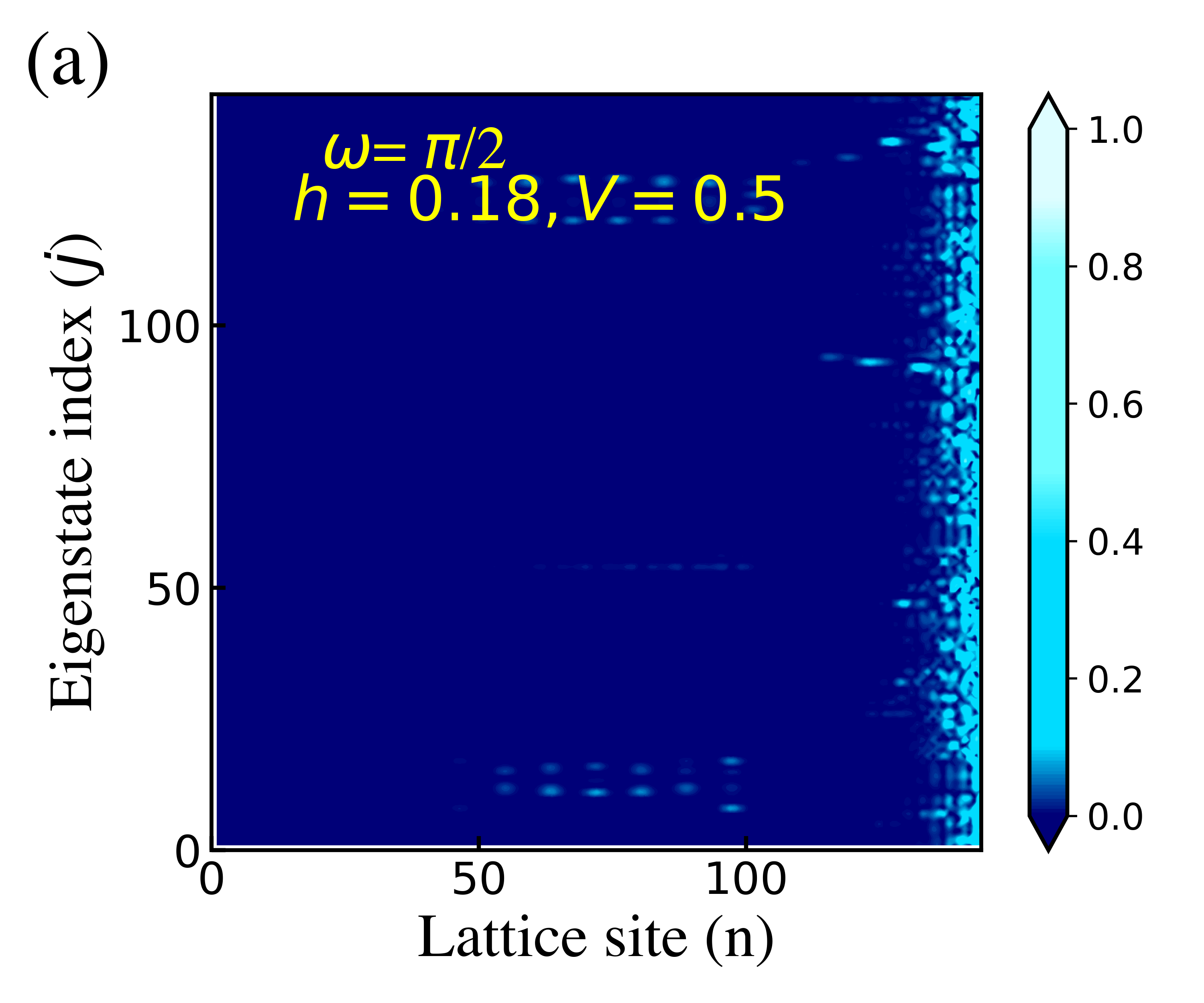}\hspace{-0.2cm}
	\includegraphics[width=0.247\textwidth,height=0.225\textwidth]{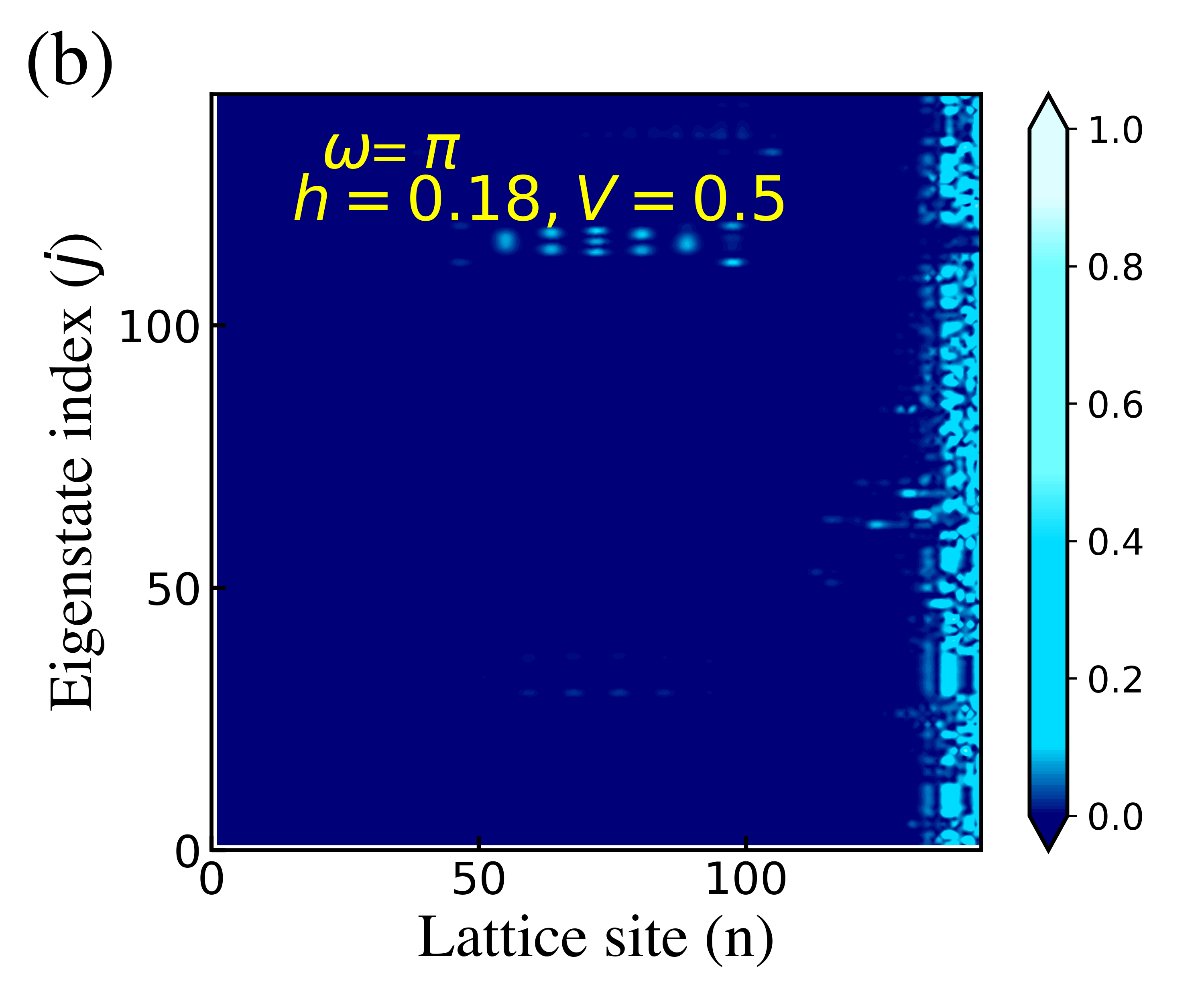}\hspace{-0.2cm}
	\includegraphics[width=0.247\textwidth,height=0.225\textwidth]{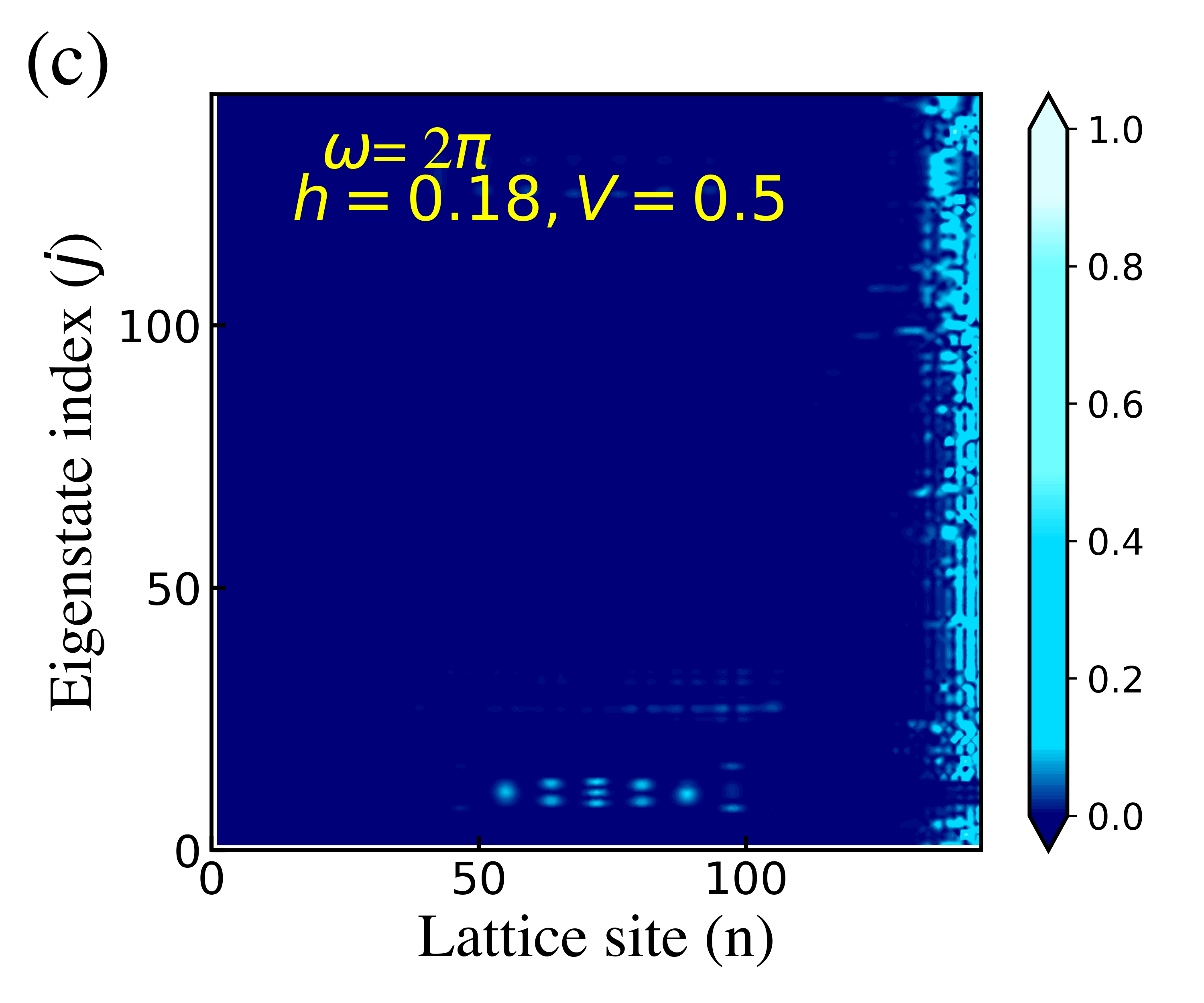}\hspace{-0.2cm}
	\includegraphics[width=0.247\textwidth,height=0.225\textwidth]{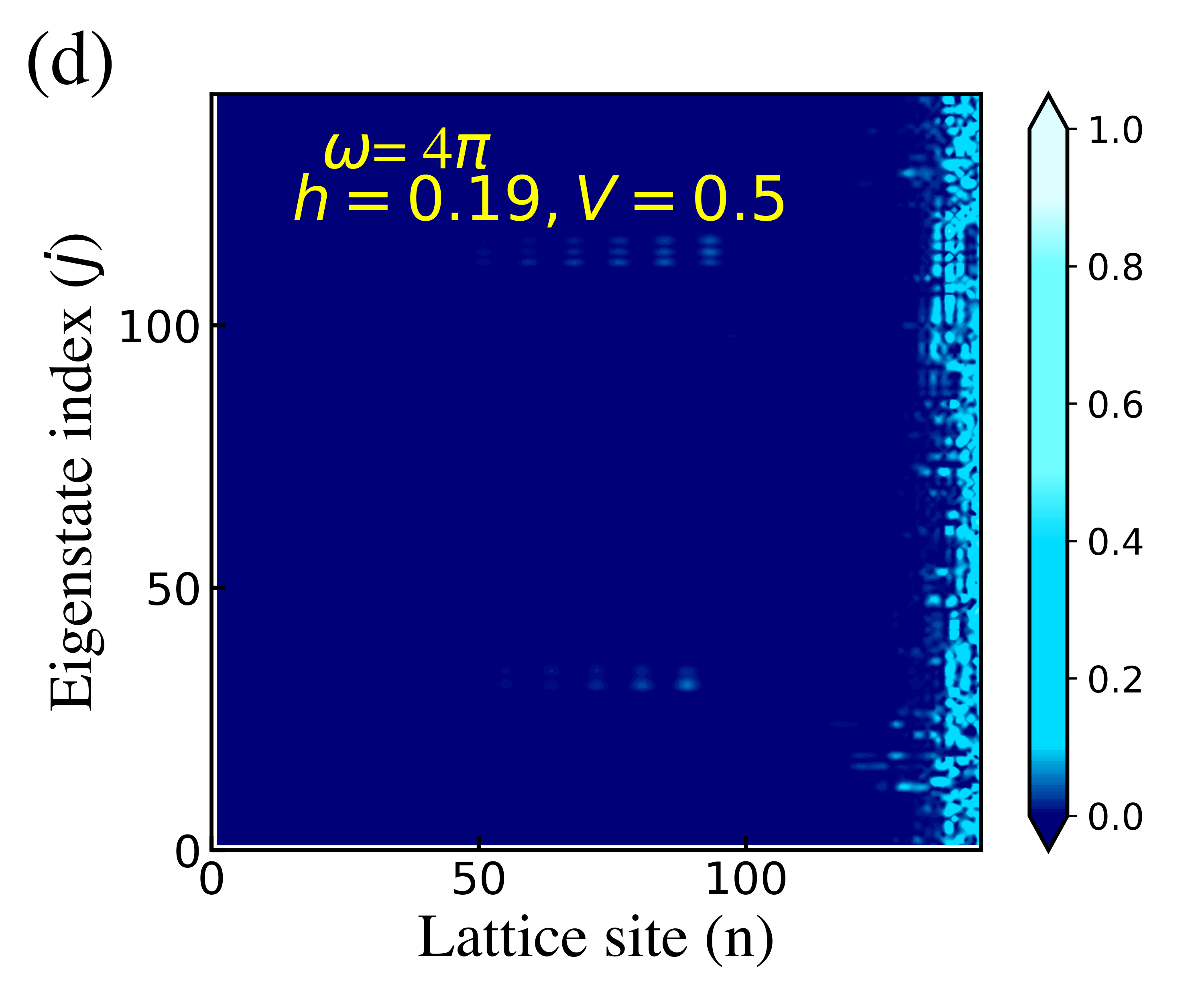}\\
	\includegraphics[width=0.247\textwidth,height=0.225\textwidth]{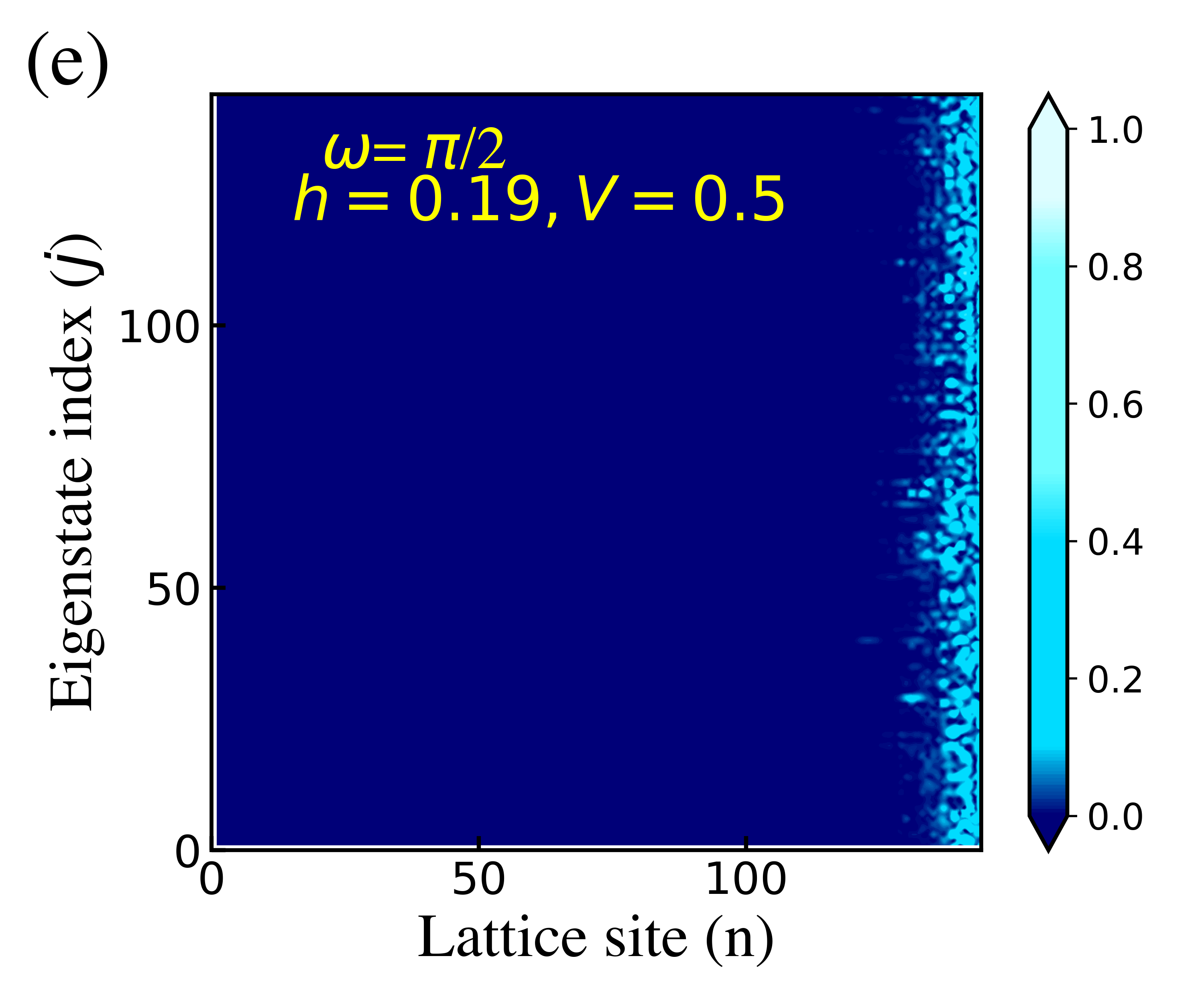}\hspace{-0.2cm}
	\includegraphics[width=0.247\textwidth,height=0.225\textwidth]{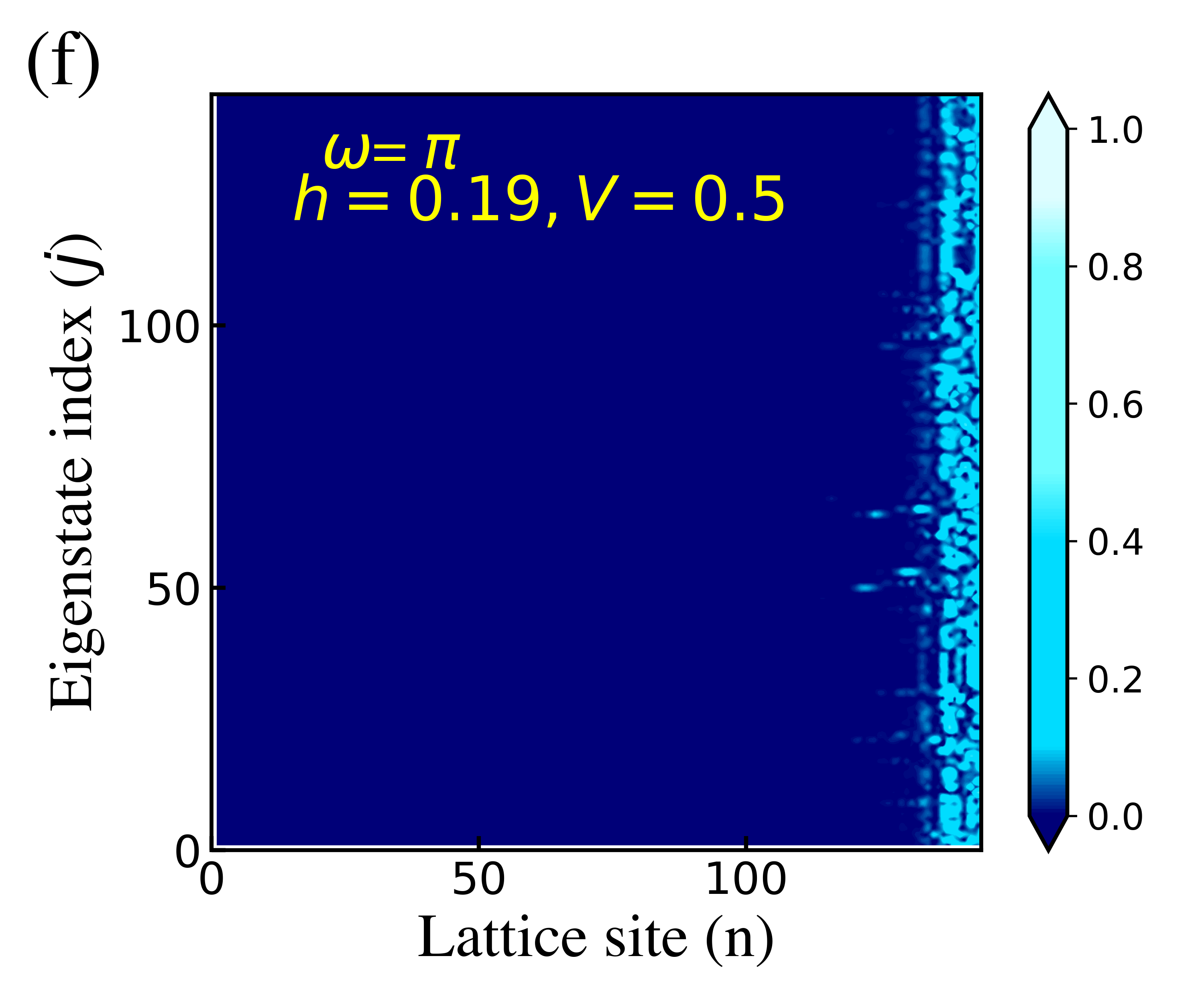}\hspace{-0.2cm}
	\includegraphics[width=0.247\textwidth,height=0.225\textwidth]{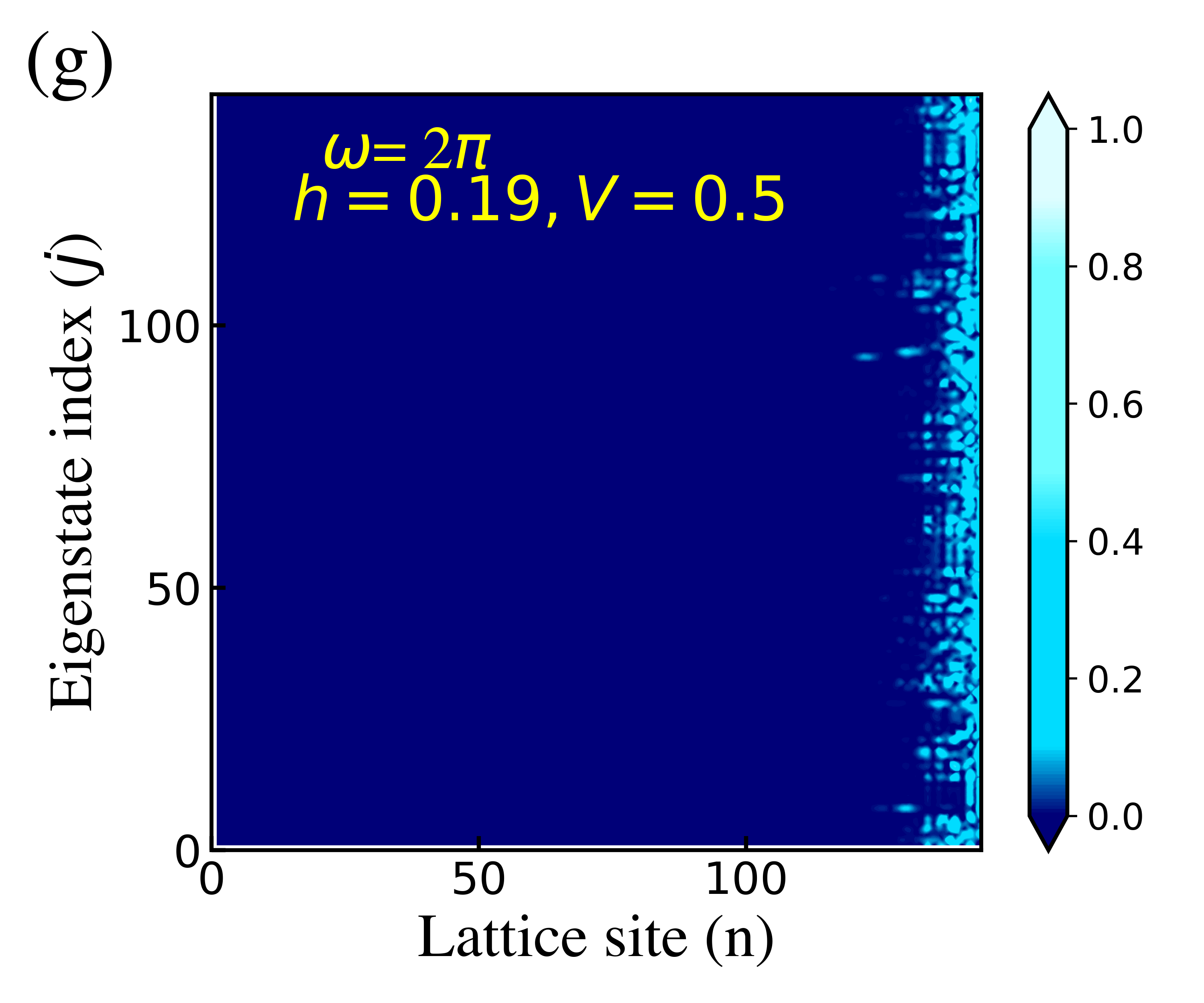}\hspace{-0.2cm}
	\includegraphics[width=0.247\textwidth,height=0.225\textwidth]{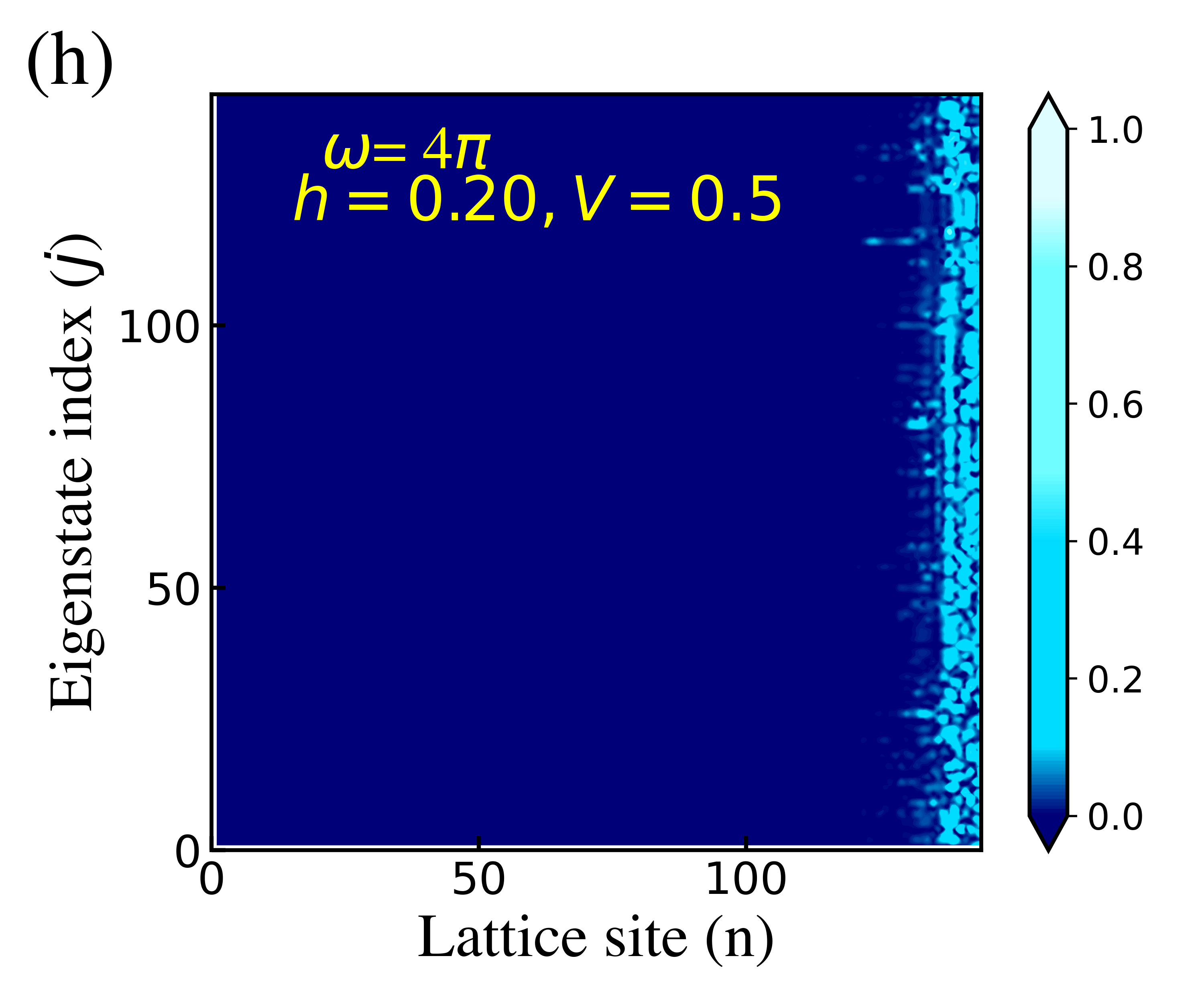}
	\caption{$|\psi|^2$ for all the eigenstates projected over the lattice sites at (a,e) $\omega=\pi/2$, (b,f) $\omega=\pi$ (c,g) $\omega=2\pi$, (d,h) $\omega=4\pi$. The figures in the upper panels correspond to the value of $h$ just before the onset of SE, whereas the ones in the lower panel correspond to the value of $h$ at which the SE appears (determined within an error of $\pm0.01$). We have considered the OBC in a lattice with 144 sites.}
	\label{Fig:Fig_S2}
\end{figure*}

In Figs.~\ref{Fig:Fig_1}(a-d) of the main text, we have illustrated the phase diagrams of the driven HN Hamiltonian considered in our work. It is well known that the delocalization-localization phase transition in the static HN systems are accompanied by a complex to real transition in the energy spectrum under the periodic boundary condition (PBC). Under the open boundary condition (OBC), such systems exhibit the skin effect (SE) as discussed in the main text. However, as we have demonstrated that in the driven systems and under the PBC, the Floquet quasienergies are always real in the delocalized regime, satisying the extended unitarity (EU) condition as shown in Figs.~\ref{Fig:Fig_2}(a,c) and Figs.~\ref{Fig:Fig_3}(a,c) in the main text for $h=0.1$ and $h=0.2$ respectively. This is in contrary to the behavior of static HN systems. Moreover, under the OBC, the Floquet quasienergy spectra changes from real (in the presence(absence) of EU(SE)) to complex (in the absence(presence) of EU(SE)). To assess the behavior of the EU in the localized regime of the phase diagram, we have presented the quasienergy spectrum under both the PBC and OBC in Figs.~\ref{Fig:Fig_S1}(a-b). It is evident that for any frequency in the drive, the Floquet quasienergies satisfy the EU condition in the localized regime of the phase diagram under both the PBC and OBC. We have verified that this EU condition persists irrespective of the strength of the quasiperiodic potential and asymmetric hopping amplitude.\\

\noindent\textbf{S.2. Demonstration of the SE in the driven HN systems}\\

To find out the critical value of $h_c$ for the onset of SE, we resort to the conventional method of determining whether all the eigenstates localize at one end of the lattice using the wave-function probabilities as demonstrated in Figs.~\ref{Fig:Fig_S2}(a-h). In an unidirectional system, all the eigenstates under the OBC pile up at one of the edges towards which there is a greater directionality of the fermionic hopping. Therefore, in our system, since the amplitude of hopping towards the right dominates over the left hopping amplitude, we say that our system possesses SE when all the states are localized towards the right end. It is clear from the upper panel that Figs.~\ref{Fig:Fig_S2}(a-d), a few eigenstates are loocalized in the bulk of the system. Therefore we do not consider them as skin modes. We have clearly illustrated the value of $h_c$ at $0.20\pm0.01$ for different driving frequencies in a lattice with 144 sites as discussed in the main text.

\end{document}